\documentclass[twocolumn]{aastex631}

\newcommand\thisplanet{HD~189733b}

\def\ut{}
\def\utn{}
\def\utnn{}

\begin{document}

\title{Eclipse Mapping with MIRI: 2D Map of \thisplanet{} from 8$\mu m$ JWST MIRI LRS Observations}

\correspondingauthor{Maura Lally}
\email{mauralally9@gmail.com, ml2289@cornell.edu}

\author[0000-0002-4443-6725]{Maura Lally}
\affiliation{Department of Astronomy and Carl Sagan Institute, Cornell University, 122 Sciences Drive, Ithaca, NY 14853, USA}

\author[0000-0002-8211-6538]{Ryan C. Challener}
\affiliation{Department of Astronomy and Carl Sagan Institute, Cornell University, 122 Sciences Drive, Ithaca, NY 14853, USA}

\author[0000-0002-8507-1304]{Nikole K. Lewis}
\affiliation{Department of Astronomy and Carl Sagan Institute, Cornell University, 122 Sciences Drive, Ithaca, NY 14853, USA}

\author[0000-0001-9164-7966]{Julie Inglis}
\affiliation{Division of Geological and Planetary Sciences, California Institute of Technology, Pasadena, CA, 91125, USA}

\author[0000-0003-3759-9080]{Tiffany Kataria}
\affiliation{Jet Propulsion Laboratory, California Institute of Technology, 4800 Oak Grove Drive, Pasadena, CA 91109, USA}

\author[0000-0002-5375-4725]{Heather A. Knutson}
\affiliation{Division of Geological and Planetary Sciences, California Institute of Technology, Pasadena, CA, 91125, USA}

\author[0000-0003-4220-600X]{Brian M. Kilpatrick}
\affiliation{National Nuclear Security Administration, U.S. Department of Energy, Washington D.C., 20585, USA}

\author[0000-0003-1240-6844]{Natasha E. Batalha}
\affiliation{NASA Ames Research Center, Moffett Field, CA 94035, USA}

\author[0000-0002-4473-0297]{Paul Bonney}
\affiliation{Jet Propulsion Laboratory, California Institute of Technology, 4800 Oak Grove Drive, Pasadena, CA 91109, USA}

\author{Ian J.\ M.\ Crossfield}
\affiliation{Department of Physics and Astronomy, University of Kansas, Lawrence, KS 66045, USA}

\author[0000-0002-6276-1361]{Trevor Foote}
\affiliation{Department of Astronomy and Carl Sagan Institute, Cornell University, 122 Sciences Drive, Ithaca, NY 14853, USA}

\author[0000-0003-4155-8513]{Gregory W. Henry}
\affiliation{Tennessee State University (retired), Nashville, TN 37216, USA}

\author[0000-0001-6050-7645]{David K. Sing}
\affil{Department of Earth \& Planetary Sciences, Johns Hopkins University, Baltimore, MD 21218, USA}
\affil{Department of Physics \& Astronomy, Johns Hopkins University, Baltimore, MD 21218, USA}

\author[0000-0002-7352-7941]{Kevin B. Stevenson}
\affiliation{Johns Hopkins Applied Physics Laboratory, Laurel, MD 20723, USA}

\author[0000-0003-4328-3867]{Hannah R. Wakeford}
\affiliation{HH Wills Physics Laboratory, University of Bristol, Tyndall Avenue, Bristol, BS8 1TL, UK}

\author[0000-0012-3245-1234]{Robert T. Zellem}
\affiliation{NASA Goddard Space Flight Center, 8800 Greenbelt Road, Greenbelt, MD 20771, USA}





\begin{abstract}


Observations and models of transiting hot Jupiter exoplanets indicate that atmospheric circulation features may cause large spatial flux contrasts across their daysides. Previous studies have mapped these spatial flux variations through inversion of secondary eclipse data. Though eclipse mapping requires high signal-to-noise data, the first successful eclipse map--made for \thisplanet{} using 8$\mu m$ Spitzer IRAC data--showed the promise of the method. JWST eclipse observations provide the requisite data quality to access the unique advantages of eclipse mapping. Using two JWST MIRI LRS eclipse observations centered on 8$\mu m$ to mimic the Spitzer bandpass used in previous studies, combined with the Spitzer IRAC 8$\mu m$ eclipses and partial phase curve (necessitated to disentangle map and systematic signals), we present a 2-dimensional dayside temperature map. Our best-fit model is a 2-component 5th-degree harmonic model with an unprecedentedly constrained eastward hotspot offset of $33.0^{+0.7}_{-0.9}$ degrees. We rule out a strong hemispheric latitudinal hotspot offset, as 3+ component maps providing latitudinal degrees of freedom are strongly disfavored. As in previous studies we find some model dependence in longitudinal hotspot offset; when we explore and combine a range of proximal models to avoid an overly constrained confidence region, we find an eastward hotspot offset of $32.5^{+3.0}_{-10.6}$ degrees, indicating the presence of a strong eastward zonal jet. Our map is consistent with some previous eclipse maps of \thisplanet{}, though it indicates a higher longitudinal offset from others. It is largely consistent with predictions from general circulation models (GCMs) at the 115 mbar level near the 8$\mu m$ photosphere. 

\end{abstract}




\section{Introduction}
\label{sec:introduction}
Hot Jupiter \thisplanet{} \citep{bouchy2005} is particularly well-suited to the characterisation of its atmospheric properties, being a relatively large planet ($R_p = 1.13 ~R_J$, \citealp{agol2010}) in a bright system ($m_V = 7.67\pm0.03$, \citealp{koen2010}) in our local solar neighborhood ($d=19.76~pc$, \citealp{gaia2023}). These factors give it a high signal-to-noise ratio (SNR), making it a favorable target for many modes of investigation. The planet has been observed using a variety of techniques, including its discovery with radial velocity, and subsequent photometric and spectroscopic observations of its transit, phase curve, and secondary eclipse \citep[e.g.,][]{deming2006, bouchy2018, moutou2020, fu2024}. Significant \textit{Spitzer} and \textit{HST} time has been dedicated to transmission spectroscopy studies of \thisplanet{}, showing the presence of water \citep{mccullough2014} and a likely detection of a haze layer in the upper atmosphere \citep{pont2008, lecavelier2008, sing2009, sing2011}. Emission spectroscopy with \textit{Spitzer} and \textit{HST} has yielded detections of water, carbon monoxide, and carbon dioxide \citep[e.g.,][]{crouzet2014, todorov2014}. Ground-based high-resolution transmission spectroscopy has also been fruitful in detecting and confirming chemical species in the planet's atmosphere, including carbon monoxide, sodium, and water \citep{redfield2008, rodler2013, birkby2013}. 

Beyond determining the chemical composition of exoplanet atmospheres, the global heat distribution across the planet is also of interest, as it aids in constraining a planet's 2D (latitude and longitude) or 3D (latitude, longitude, and altitude) thermal structure. Hot Jupiters provide an ideal laboratory for studying this thermal structure: in general, they are tidally locked and therefore highly irradiated on their permanent daysides, providing extreme climates with stark day/night contrasts and intensified atmospheric circulation features. General Circulation Models (GCMs) have played an important role in constraining the atmospheric dynamics of hot Jupiters, either by exploring the general hot Jupiter parameter space or modeling a specific observed hot Jupiter (see \citealp{showman2020} for a review). Taken together, the combination of observations and GCMs show that jets and vortices in a hot Jupiter atmosphere can cause large spatial flux contrasts across the daysides of hot Jupiters from their equators to their poles. With the advent of JWST, it is now possible to test those predictions for many hot Jupiters using maps derived from observational data.

Although phase curve mapping and eclipse mapping are both successful methods to create longitudinal heat maps of exoplanets, they have unique challenges and benefits. In general, for data of the same quality, a planet’s longitude of the hemisphere of maximum brightness can be located more effectively with phase curve mapping than eclipse mapping, simply by virtue of eclipse ingress and egress duration being so much shorter than the duration of the total orbital period \citep{majeau2012}. This also means that eclipse mapping requires a higher baseline data quality than does phase curve mapping \citep{challener2022, schlawin2023}. That being said, when possible, eclipse mapping has unique advantages: unlike phase curve observations, eclipses of inclined orbits can yield information about latitudinal flux distributions \citep{majeau2012, dewit2012}. Eclipse mapping is also less sensitive to some sources of variability which occur on the scale of a planetary orbit, including stellar variability \citep{schlawin2023}. As a practical matter, although eclipse mapping requires higher quality data, it can also be used to find the longitude of the hemisphere of maximum brightness in a fraction of the telescope time that we need to find the same information from phase curve mapping. With data of the requisite quality, eclipse mapping can provide much higher spatial resolution in longitude and allows measurement of much smaller-scale features \citep{rauscher2018, challener2022}.  

\thisplanet{} has been studied using phase curves and eclipse mapping, both separately and simultaneously \citep{knutson2007, agol2010, majeau2012, dewit2012, rauscher2018, challener2022}, making it an interesting object for follow-up mapping efforts with updated and higher-SNR eclipse data. Phase curve and eclipse observations of \thisplanet{}, obtained using the $8~\mu m$ channel of the Spitzer IRAC instrument \citep{fazio2004}, enabled longitudinal temperature maps of the planet’s dayside \citep{cowan&agol2008, majeau2012,dewit2012}. The first of these longitudinal maps of \thisplanet{} was created by inverting a phase-curve observation of about half of a planetary orbit. Additional transits and secondary eclipses were observed \citep{agol2010}, eventually leading to the first successful eclipse maps of an exoplanet \citep{majeau2012,dewit2012}. Favorability of \thisplanet{} for follow-up JWST eclipse mapping was assessed in \citet{boone2024}, which calculated an eclipse mapping metric (EMM) for a sample of hot Jupiter exoplanets based on their planet to star flux ratio, the duration of eclipse ingress and egress, and the magnitude of their host star. According to this EMM, \thisplanet{} remains the most favorable target for mapping with MIRI LRS. 

Previous eclipse mapping studies have consistently suggested an eastward hotspot offset for \thisplanet{}, suggesting a high degree of atmospheric circulation. However, these previous studies found hotspot offset values to be highly uncertain and model dependent \citep{majeau2012, dewit2012}. This research presents a new 2D map of \thisplanet{}’s dayside atmosphere centered at $8\mu m$, in an attempt to compare and update maps produced with Spitzer’s $8\mu m$ band. 
\ut{Although using the full MIRI bandpass would provide the highest SNR, it effectively integrates over several pressure levels, and therefore the resulting maps would not be comparable to previous Spitzer maps. We leave study of the wavelength dependence of the temperature map of \thisplanet\ to future work.}
JWST MIRI LRS data offers several advantages over data from Spitzer: \ut{even binned to match the Spitzer $8 \mu m$ bandpass,} the data has a much higher SNR, and the instrumental ramp effects are much less severe. These factors will allow detection of much smaller scale structure in the ingress and egress. In turn, dayside temperature maps will be much better constrained.

In Section \ref{sec:methods}, we discuss the data collection, reduction and analysis used to create a 2D $8\mu m$ temperature map of \thisplanet{}. In Section \ref{sec:analysis tests} we discuss the sensitivity of our eclipse maps to JWST MIRI LRS systematics and how we approach degeneracies. In Section \ref{sec:results} we present our results. In Section \ref{sec:discussion} we discuss our results in further detail and compare them to previous eclipse maps of the planet. In Section \ref{sec:conclusion} we conclude and discuss potential for future work.

\section{Methods} 
\label{sec:methods}

\begin{figure}[ht!]
\centering
   \includegraphics[width=3.25in]{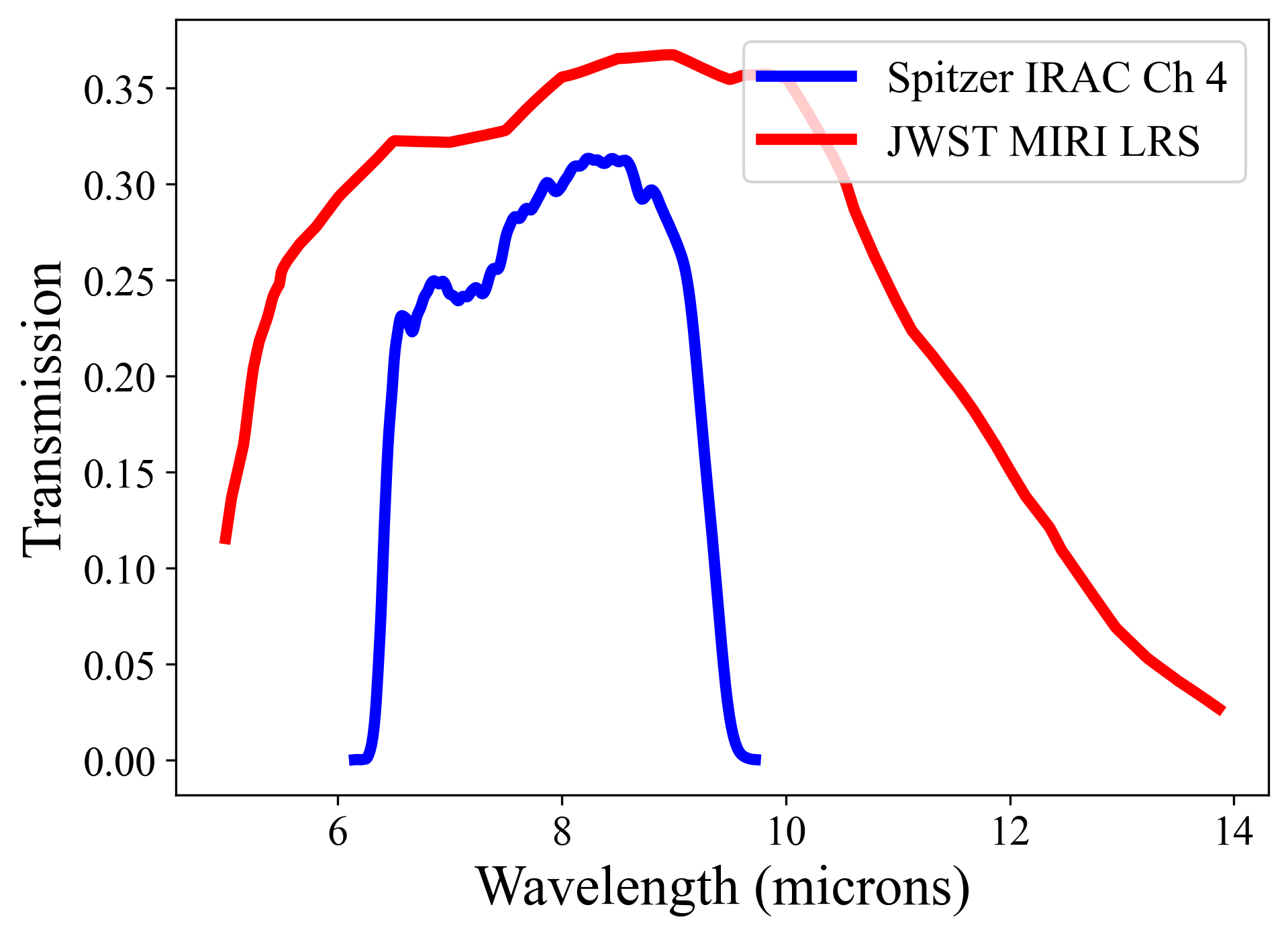}
   \caption{Throughputs of MIRI LRS and Spitzer IRAC $8 \mu m$. Our observations were taken over the full MIRI bandpass, and then weighted by the Spitzer IRAC $8\mu m$ bandpass to create an $8\mu m$ eclipse map for comparison to previously published Spitzer results \citep{majeau2012, dewit2012}.}
   \label{fig:throughputs}
\end{figure}

In order to create a map of dayside spatial flux variations on \thisplanet{}, we first acquire and reduce two approximately $4.5$ hour eclipse observations from MIRI LRS. We use a subsection of the MIRI LRS bandpass to create lightcurves centered on $8\mu m$, weighted according to the Spitzer IRAC $8\mu m$ bandpass (shown in Figure \ref{fig:throughputs}). This Spitzer IRAC bandpass covers a section of the contribution function of \thisplanet{} which corresponds to about the ~100 mbar level of the atmosphere \citep[e.g.,][]{showman2009}. We then use \texttt{ThERESA}\footnote{\texttt{ThERESA} documentation: https://theresa.readthedocs.io/en/latest/} \citep{challener2022} to fit our $8\mu m$ eclipse lightcurves, and create a temperature map of the planet’s dayside. We then compare our resulting lightcurve fits and temperature maps to those created with Spitzer IRAC $8\mu m$ bandpass eclipse data. 

\subsection{Data Acquisition}
\label{sec:data acquisition}
We use JWST MIRI LRS to observe two separate eclipses of \thisplanet{} (GO program 2021, PI Brian Kilpatrick). The first observation ran October 18, 2022 16:38:42 UT to Oct 18, 2022 22:29:04 UT; the second observation ran June 30, 2023 19:49:47 UT to July 1 2023 01:41:22 UT. Both visits used the SLITLESSPRISM subarray for observations. We obtained about 4.5 hours of usable data on each visit, with each eclipse including a total of 17,035 integrations of 0.95 seconds each \citep{inglis2024}.

\subsection{Data Reduction}
\label{sec:data reduction}
We use data reduced by \citet{inglis2024} using the \texttt{Eureka!} \citep{bell2022} pipeline, version 0.9. For Stages 1 and 2 of the reduction, \texttt{Eureka!} mostly relied on the default \texttt{JWST Science Calibration Pipeline} (\texttt{JWST}) settings to perform ramp fitting, non-linearity, gain scaling, saturation and data quality flagging, flat fielding, and source type steps. For background subtraction and flux calibration, \texttt{JWST} pipeline Stage 2 steps were skipped in favor of \texttt{Eureka!} custom steps. This choice was made to avoid unnecessary noise being added to the final lightcurves. Stage 3 of \texttt{Eureka!} then created a flux curve centered on  $8\mu m$ for each eclipse. Pixels flagged in the data quality array, as well as ``NaN" and ``inf" pixels, were masked. Background subtraction at $8 \mu m$ was performed by fitting a linear function in the cross-dispersion direction, using only pixels outside of a 24-pixel aperture centered on the trace. Both optimal and standard box extraction were tested, with negligible difference on a resulting MIRI emission spectrum; optimal extraction was therefore used for the final analysis, with the gain manually set to 3.1 \citep{bell2022}. Together, the light curves have a standard deviation of the normalized residuals (SDNR) of 334 ppm. For more details regarding the data reduction, see \citet{inglis2024}.

In keeping with other JWST time-series analyses \citep[e.g.,][]{CO2_ERS,kempton2023,Grant2023}, we use a secondary reduction of the same data by \citet{inglis2024} using the \texttt{SPARTA} \citep{kempton2023} pipeline to ensure that our result is robust to different reduction choices. In lieu of the \texttt{JWST} Stage 1 and Stage 2 used by \texttt{Eureka!}, the \texttt{SPARTA} pipeline uses a custom slope-fitting routine for the ramps, which accounts for non-linear group response. It also uses a custom cosmic ray removal step. These different treatments of instrumental effects such as switch charge decay and last-frame effect, as well as physical outliers such as cosmic rays, provide a fully independent data reduction which we can test to validate our results from the \texttt{Eureka!} reduction. After the custom Stage 1 steps, \texttt{SPARTA} executes similar steps to the \texttt{JWST} Stage 2 reduction, such as flat fielding, gain correction (also set to a constant 3.1 as in the \texttt{Eureka!} reduction, see \citet{bell2022}), pixel masking, and dark subtraction. Background subtraction is performed with a row-wise median subtraction from windows above and below the trace. Optimal and standard extraction from the trace was done similarly to the \texttt{Eureka!} reduction, with a window with half-width of 5-pixels centered on the trace. As with the \texttt{Eureka!} reduction, optimal extraction is adopted for best comparison. 

\subsection{Data Analysis}
\label{sec:data analysis}

To fit our two eclipse lightcurves and create a map of the planet’s dayside, we use the eclipse-mapping code \texttt{ThERESA} \citep{challener2022}. In general, eclipse mapping relies on a visibility function to determine the visible longitudes of the planet during eclipse. The planet visibility reduces due to both the planet's rotation through an angle with respect to the observer, and due to the eclipse by the star, formalized as $V = LS$ where $V$ is the visibility, $L$ is the line-of-sight visibility, and $S$ is eclipse visibility. The planet’s true latitude and longitude remain fixed, but as the planet rotates, the “sub-observer” point ($\theta=0, \phi=0$) moves with regard to latitude and longitude. The stellar visibility ($S$) term is what enables eclipse mapping: as the planet moves behind the star and re-emerges, different regions on the planet are visible at different times, so emission can be attributed to individual regions instead of integrated hemispheres, offering a much more detailed map than is possible from other methods.

The purpose of using \texttt{ThERESA} in this analysis is its unique approach to fitting light curves using orthogonalized spherical harmonic maps to create a dayside temperature map \citep{starry}. The code uses principal component analysis of eclipse lightcurves generated from spherical harmonic maps to determine a set of orthogonal light curves, called eigencurves \citep{rauscher2018}. A least-squares routine identifies an initial guess for a best-fit linear combination of eigencurves, which is converted into a temperature map. The Markov-chain Monte Carlo implementation \texttt{mc3} \citep{cubillos2017} is used to explore the parameter space and estimate uncertainties. The maximum allowed complexity of the spherical harmonic maps used in the fit is a user-defined parameter, $l_{max}$. In our analysis, we set $l_{max} = 10$; this includes higher complexity than previous eclipse mapping efforts, in an attempt to explore a large range of models and ensure the optimal fit. This is further discussed in Section \ref{sec:discussion}. 

The astrophysical model’s functional form is

\begin{equation} \label{eq:sys}
F_{sys}(t) = 1 + F_j(c_0 Y_0^0 + \sum_{i=1}^N c_i E_i)
\end{equation}
where $F_{sys}$ is the system flux, $F_{j}$ is a fitted parameter for the $j^{th}$ eclipse (when fitting multiple eclipses, one remains fixed and the rest are rescaled using $F_j$ to account for stellar activity), $N$ is the number of eigencurves used, $c_i$ are the eigencurve weights, $E_i$ are the eigencurves, $Y_0^0$ is the uniform-map spherical harmonic component, and $F_{star}$ is the light curve of the star (unity, because the curves are normalized). Like $l_{max}$, the parameter $N$ has a user-defined maximum; eigencurves within the $l_{max}$ limit are ranked by their variance (how strongly observable they are in the data), and only the $N$ highest-variance eigencurves are used in the final map, a process which is optimized by minimizing the Bayesian information criterion (BIC) \citep{schwarz1978, liddle2007}. \ut{This model selection criteria is consistent with past mapping work \citep[e.g.][]{challener_rauscher2023, schlawin2023, challener2024, schlawin2024}.}  

\texttt{ThERESA} creates a measured brightness map and then converts it to a temperature map. The conversion assumes a PHOENIX model stellar spectrum \citep{husser2013} corresponding to the stellar parameters from \cite{rosenthal_california_2021} (effective temperature $T_{eff}$ = 5012 K, surface gravity log(g)= 4.5, and metallicity [Fe/H] = 0.04), which is integrated over the throughput wavelengths (in this case, about $6.5\mu m$ to $9.5 \mu m$, weighted to match the Spitzer $8\mu m$ throughput; see Figure \ref{fig:throughputs}). The planet is assumed to be a blackbody, and Planck functions over a range of possible planetary temperatures are integrated over the throughput wavelengths to create a sampling of $T_p(F_p)$. This relationship is then interpolated to the measured planet-to-star flux ratio to determine a temperature. 

\begin{deluxetable}{ccc}[htbp]
\label{table:params_table}
\tablecaption{Fixed Input Parameters for \textsc{ThERESA} Model}
\tablehead{
  \colhead{Parameter} & 
  \colhead{Units}     &
  \colhead{Value}
}
\startdata
$M_s$ & Stellar mass $(M_\odot)$$^a$ & $0.807 \pm0.005$ \\
$R_s$ & Stellar radius $(R_\odot)$$^b$ & $0.752\pm0.025$ \\
$M_p$ & Planet mass $(M_\odot)$$^b$ & $0.00110\pm4\text{e-}5$ \\
$R_p$ & Planet radius $(R_\odot)$$^a$ & $0.11680\pm0.00014$ \\
$p_o$ & Planet orb. period (d)$^c$$^e$ & $2.218575\pm 6\text{e-}8$ \\
$p_r$ & Planet rot. period (d)$^c$ & $2.218575\pm 6\text{e-}8$ \\
$e$ & Eccentricity$^e$ & $0$ \\
$i$ & Inclination (deg)$^c$ & $85.71\pm0.024$ \\
$a$ & Semi-major axis (AU)$^a$ & $0.03099\pm7\text{e-}5$ \\
$t_t$ & Transit time (d, UTC)$^d$$^e$ & $59838.6863112\pm4\text{e-}6$ \\
\enddata
\footnotesize{$^a$ Calculated using values from \citet{agol2010} and $R_s$ from \citet{southworth2010}. \\$^b$\citet{southworth2010}. \\$^c$\citet{agol2010}. \\$^d$ Determined by fitting MIRI LRS transits, \citet{transit_prop2021}.}
\\$^e$ Fixing the orbital period, transit time, and a zero eccentricity also effectively fixes the mid-eclipse time for each of our fits. 
\end{deluxetable}

The MIRI instrumental ramp is strongest at the beginning of the observation, and affects each eclipse observation differently. The \texttt{ThERESA} systematic modeling process was optimized for this analysis to better fit planets for which systematics differ over each observation. To minimize the effect of the ramp while also properly characterizing it for removal, we remove the first 30 minutes of data from the first eclipse and the first hour of data from the second eclipse. To further treat the instrumental ramp, we use an exponential ramp function of form
\begin{equation} \label{eq:ramp}
    R = b + p_1 t + Ae^{\frac{-t}{\tau}} 
\end{equation}
where $t$ is the `local time' (starting at the beginning of each eclipse observation, defined prior to cutting the initial 500 integrations), and $b$ is a constant offset to accommodate errors in stellar normalization. We found that this parameterization worked well to characterize and remove the instrumental ramp from the first eclipse, while the ramp's effect on the second observed eclipse was well characterized by a simple linear model $F = b + p_1 t$, effectively setting $A=0$. We tested several other ramp functions, including a combination of two exponential terms, as used in Stage 5 of the \texttt{Eureka!} pipeline \citep{Eureka}. We prefer the linear + exponential ramp in Equation \ref{eq:ramp} because it was strongly preferred by our BIC model selection and has been used successfully in recent similar studies \citep{hammond2024}.

In addition to an instrumental ramp, we also perform positional detrending to reduce the red noise present in our residuals. We use the functional form
\begin{equation} \label{eq:pos}
    S = 1 + d_x x_{pos} + d_y y_{pos} + d_w w_{PSF} 
\end{equation}
to take into account shifts of the trace on the detector, where $x_{pos}$ is the x-position, $y_{pos}$ is the y-position, and $w_{PSF}$ is the width of the point spread function. We fit for coefficients $d_x, d_y$ and $d_w$.

Our ramp and positional detrending are co-fit simultaneously with the eclipse model to prevent our analysis from attributing ramp effects to true eclipse signal, in accordance with the suggestions from \citet{schlawin2023} on disentangling long-term baseline instrumental drifts from eclipse map models. This method has also been performed in previous eclipse mapping studies \citep{challener_rauscher2023}. Since the two input observations are the same target in the same bandpass at different times, we fit the same astrophysical parameters to both visits, but allow ramp and positional detrending parameters to differ between visits. We do not apply ramp or positional detrending to the Spitzer eclipses and partial phase curve used in this analysis as this has already been performed \citep{agol2010}. We do co-fit that data with our eclipse model. Our full fit model is then parameterized as 

\begin{equation} \label{eq:full_model}
    M = F_{sys} * R * S
\end{equation}

where $F_{sys}, R$ and $S$ are defined in Equations \ref{eq:sys}, \ref{eq:pos} and \ref{eq:ramp}.

\texttt{ThERESA} requires system parameter inputs to calculate the eigencurves $E_i$. The input parameters used in this analysis appear in Table \ref{table:params_table} \citep{agol2010, southworth2010, inglis2024}.

\section{Sensitivity of Maps to Systematics}
\label{sec:analysis tests}

We performed a number of sensitivity analyses before reaching a robust result as presented in Section \ref{sec:results}. These are described in detail in the sections below. 

\subsection{Individual Eclipse Fits}
\label{sec:individual eclipse fits}

We first attempt \texttt{ThERESA} mapping on the MIRI eclipses separately, to investigate the certainty of a result from a single JWST eclipse. We run \texttt{ThERESA} in 2D mode, with 50,000 samples and a burn-in of 25,000 to ensure chain convergence. Our MCMC explores parameters space for eigencurve components up to the defined complexity limit (($l_{max}=6, N=6$), the ramp parameters ($b,~p_1,~A$ and $\tau$), and the positional detrending parameters ($d_x,~d_y$ and $d_w$). Latitude and longitude cell resolution of the output map is user-defined and, in this analysis, set to an arbitrary $12$ by $24$. This parameter has a minimal impact on the resulting fit; it only sets the visual resolution of the output map and the resolution at which \texttt{ThERESA} enforces a positive dayside flux condition.

\begin{figure*}[ht!]
\centering
	\includegraphics[width=6.25in]{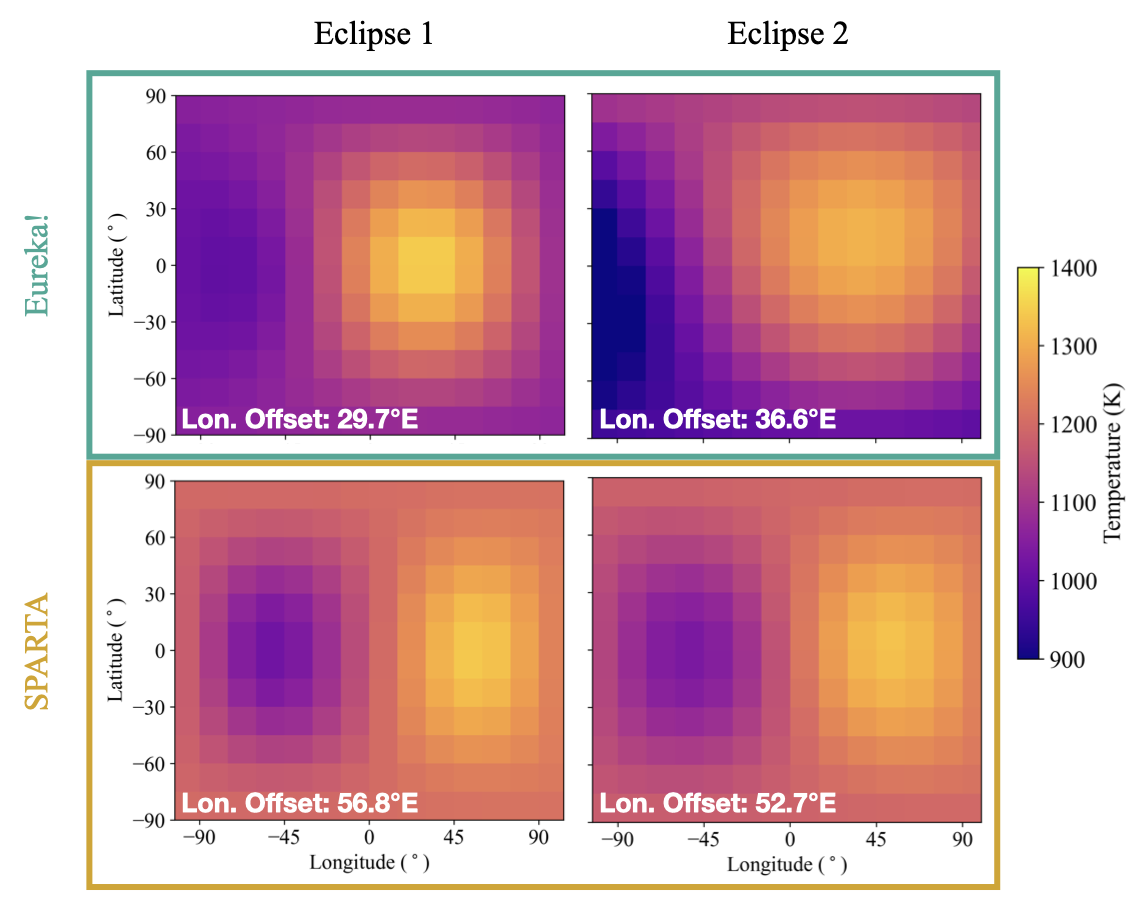}
    \caption{Dayside maps of HD189733b fit to single individual eclipses. The top row shows maps fit to the \texttt{Eureka!} reduction, while the bottom row shows maps fit to the \texttt{SPARTA} reduction. Shown in each map is the resulting longitudinal hotspot offset; although we achieve high-precision maps, when we fit them individually, they are not necessarily consistent across different eclipses nor across different data reduction methods.
    }
    \label{fig:maps_individual}
\end{figure*}

We perform this analysis on both the eclipses individually, from both the \texttt{Eureka!} reduction as well as the \texttt{SPARTA} reduction, yielding four maps, shown in Figure \ref{fig:maps_individual}. Our preferred models, priors, and posteriors are shown in Table \ref{table:individual_fits_table}. As shown in the Figure and Table, although single eclipses were impressively constraining (especially the first eclipse), the four resulting maps yielded significantly differing results for the hotspot offset. Differences were especially large between the \texttt{Eureka!} and \texttt{SPARTA} reductions. 

\startlongtable
\begin{deluxetable*}{cccc}
\label{table:individual_fits_table}
\tablecaption{Analysis Test Posteriors}
\tablehead{
  \colhead{Parameter}   &
  \colhead{Prior}   &
  \colhead{Eureka Post.} &
  \colhead{Sparta Post.}
}
\startdata
\hline
\multicolumn{4}{l}{Eclipse 1 Individual Fits} \\
\multicolumn{4}{l}{(\texttt{Eureka!} model: $l=3, N=2$; \texttt{SPARTA} model: $l=5, N=2$)} \\
\hline
$C_0$   & $[-1,1]$  &  $0.00260^{+0.00012}_{-0.00018}$      & $0.00320^{+0.00013}_{-0.00024}$ \\
$C_{1}$ & $[-1,1]$  &  $0.000321^{+0.000091}_{-0.000067}$   & $-0.000028^{+0.000122}_{-0.000071}$ \\
$C_{2}$ & $[-1,1]$  &  $0.000222^{+0.000028}_{-0.000017}$   & $0.000240^{+0.000017}_{-0.000017}$ \\
$d_x$   & $[-\infty,\infty]$ & $-0.00113^{+0.00031}_{-0.00028}$ & $-0.01757^{+0.00056}_{-0.00052}$ \\
$d_y$  &  $[-\infty,\infty]$ & $-0.00033^{+0.00066}_{-0.00067}$ & $-0.00107^{+0.00071}_{-0.00105}$ \\
$d_w$  & $[-\infty,\infty]$  & $-0.0297^{+0.0125}_{-0.0058}$    & $-0.0495^{+0.0054}_{-0.0060}$ \\
$b$  & $[0.8,1.2]$  & $0.997203^{+0.000014}_{-0.000023}$  & $0.996961^{+0.000036}_{-0.000064}$ \\
$m$  & $[-1,1]$     & $-0.00152^{+0.00036}_{-0.00017}$    & $0.00004^{+0.00051}_{-0.00032}$ \\
$A$  & $[-0.01,0.01]$ & $0.000213^{+0.000021}_{-0.000022}$  & $0.000444^{+0.000086}_{-0.000049}$ \\
$\tau$  & $[0.0,0.2]$ & $0.01026^{+0.00040}_{-0.00076}$     & $0.04241^{+0.00061}_{-0.00056}$ \\ 
Lon. Offset & ... & $29.71^{+0.53}_{-5.06}$ & $56.81^{+5.31}_{-7.14}$ \\
\hline
\multicolumn{4}{l}{Eclipse 2 Individual Fits} \\
\multicolumn{4}{l}{(\texttt{Eureka!} model: $l=1, N=3$; \texttt{SPARTA} model: $l=2, N=2$)} \\
\hline
$C_0$   & $[-1,1]$ & $0.00257^{+0.00066}_{-0.00027}$ & $0.003167^{+0.000110}_{-0.000096}$ \\
$C_{1}$ & $[-1,1]$ & $0.00036^{+0.00016}_{-0.00040}$ & $-0.000013^{+0.000052}_{-0.000059}$ \\
$C_{2}$ & $[-1,1]$ & $0.000339^{+0.000036}_{-0.000089}$ & $0.000245^{+0.000016}_{-0.000014}$ \\
$C_{3}$ & $[-1,1]$ & $0.000098^{+0.000039}_{-0.000078}$ & ... \\
$d_x$   & $[-\infty,\infty]$ & $-0.00105^{+0.00034}_{-0.00034}$ & $-0.01728^{+0.00051}_{-0.00048}$ \\
$d_y$   & $[-\infty,\infty]$ & $-0.00148^{+0.00065}_{-0.00076}$ & $-0.00017^{+0.00079}_{-0.00078}$ \\
$d_w$   & $[-\infty,\infty]$ & $-0.0320^{+0.0056}_{-0.0050}$    & $-0.0536^{+0.0053}_{-0.0050}$ \\
$b$     & $[0.8,1.2]$   & $0.997047^{+0.000023}_{-0.000013}$ & $0.9969463^{+0.0000085}_{-0.0000076}$ \\
$m$     & $[-1,1]$      & $0.00093^{+0.00019}_{-0.00046}$ & $0.00203^{+0.00011}_{-0.00013}$ \\
Lon. Offset & ... & $36.60^{+14.25}_{-12.97}$ & $52.71^{+4.93}_{-6.37}$ \\
\hline
\multicolumn{4}{l}{Joint MIRI Eclipse Fits} \\
\multicolumn{4}{l}{(\texttt{Eureka!} model: $l=5, N=2$; \texttt{SPARTA} model: $l=5, N=2$)} \\
\hline
$C_0$   & $[-1,1]$ & $0.002799^{+0.000078}_{-0.000077}$ & $0.003171^{+0.000086}_{-0.000086}$ \\
$C_{1}$ & $[-1,1]$ & $0.000205^{+0.000042}_{-0.000042}$ & $-0.000024^{+0.000046}_{-0.000045}$ \\
$C_{2}$ & $[-1,1]$ & $0.0002075^{+0.0000099}_{-0.0000098}$ & $0.000230^{+0.000011}_{-0.000010}$ \\
$N_1$  &  $[0.8,1.2]$ & $1.0$ & $1.0$ \\
$d_{x,1}$  & $[-\infty,\infty]$ & $-0.00115^{+0.00028}_{-0.00029}$ & $-0.01759^{+0.00046}_{-0.00045}$ \\
$d_{y,1}$  & $[-\infty,\infty]$ & $-0.00044^{+0.00053}_{-0.00053}$ & $-0.00111^{+0.00074}_{-0.00075}$ \\
$d_{w,1}$  & $[-\infty,\infty]$ & $-0.0329^{+0.0040}_{-0.0041}$ & $-0.0496^{+0.0049}_{-0.0049}$ \\
$b_1$  & $[0.8,1.2]$ & $0.9972163^{+0.0000090}_{-0.0000103}$ & $0.99688^{+0.00011}_{-0.00019}$ \\
$m_1$  & $[-1,1]$ & $-0.00173^{+0.00012}_{-0.00011}$ & $0.00051^{+0.00098}_{-0.00065}$ \\
$A_1$  & $[-0.01,0.01]$ & $0.000203^{+0.000020}_{-0.000019}$ & $0.00053^{+0.00019}_{-0.00011}$ \\
$\tau_1$  & $[0.0,0.2]$ & $0.00925^{+0.00181}_{-0.00096}$ & $0.053^{+0.017}_{-0.011}$ \\
$N_2$  &  $[0.8,1.2]$ & $0.9978^{+0.0028}_{-0.0027}$ & $1.0096^{+0.0038}_{-0.0039}$ \\
$d_{x,2}$  & $[-\infty,\infty]$ & $-0.00112^{+0.00032}_{-0.00031}$ & $-0.01730^{+0.00052}_{-0.00053}$ \\
$d_{y,2}$  & $[-\infty,\infty]$ & $-0.00117^{+0.00056}_{-0.00054}$ & $-0.00016^{+0.00083}_{-0.00078}$ \\
$d_{w,2}$  & $[-\infty,\infty]$ & $-0.0329^{+0.0042}_{-0.0042}$ & $-0.0535^{+0.0051}_{-0.0050}$ \\
$b_2$  & $[0.8,1.2]$ & $0.9970676^{+0.0000066}_{-0.0000069}$ & $0.9969508^{+0.0000073}_{-0.0000072}$ \\
$m_2$  & $[-1,1]$ & $0.000701^{+0.000095}_{-0.000092}$ & $0.00197^{+0.00010}_{-0.00010}$ \\
Lon. Offset & ... & $39.44^{+2.30}_{-2.51}$ & $54.01^{+3.06}_{-3.16}$ \\
\hline
\multicolumn{4}{l}{Joint MIRI + Spitzer Eclipse and Partial Phase Curve Fits} \\
\multicolumn{4}{l}{(\texttt{Eureka!} model: $l=5, N=2$; \texttt{SPARTA} model: $l=5, N=2$)} \\
\hline
$C_0$   & $[-1,1]$ & $0.002554^{+0.000032}_{-0.000032}$ & $0.002534^{+0.000028}_{-0.000028}$ \\
$C_{1}$ & $[-1,1]$ & $0.000334^{+0.000014}_{-0.000014}$ & $0.000321^{+0.000015}_{-0.000014}$ \\
$C_{2}$ & $[-1,1]$ & $0.0002025^{+0.0000089}_{-0.0000091}$ & $0.0002183^{+0.0000094}_{-0.0000094}$ \\
$N_1$  & $[0.8,1.2]$ & $1.0$ & $1.0$ \\
$d_{x,1}$  & $[-\infty,\infty]$ & $-0.00116^{+0.00028}_{-0.00029}$ & $-0.01752^{+0.00044}_{-0.00045}$ \\
$d_{y,1}$  & $[-\infty,\infty]$ & $-0.00048^{+0.00054}_{-0.00053}$ & $-0.00101^{+0.00075}_{-0.00074}$ \\
$d_{w,1}$  & $[-\infty,\infty]$ & $-0.0332^{+0.0040}_{-0.0040}$ & $-0.0504^{+0.0050}_{-0.0050}$ \\
$b_1$  & $[0.8,1.2]$ & $0.997208^{+0.000016}_{-0.000199}$ & $0.99654^{+0.00022}_{-0.00040}$ \\
$m_1$  & $[-1,1]$ & $-0.00160^{+0.00142}_{-0.00019}$ & $0.0023^{+0.0016}_{-0.0011}$ \\
$A_1$  & $[-0.01,0.01]$ & $0.000221^{+0.000171}_{-0.000025}$ & $0.00091^{+0.00040}_{-0.00023}$ \\
$\tau_1$  &  $[0.0,0.2]$ & $0.0133^{+0.0385}_{-0.0044}$ & $0.076^{+0.026}_{-0.017}$ \\
$N_2$  & $[0.8,1.2]$ & $1.0003^{+0.0087}_{-0.0040}$ & $1.0147^{+0.0036}_{-0.0037}$ \\
$d_{x,2}$  & $[-\infty,\infty]$ & $-0.00113^{+0.00031}_{-0.00032}$ & $-0.01740^{+0.00051}_{-0.00050}$ \\
$d_{y,2}$  & $[-\infty,\infty]$ & $-0.00113^{+0.00055}_{-0.00056}$ & $-0.00006^{+0.00080}_{-0.00079}$ \\
$d_{w,2}$  & $[-\infty,\infty]$ & $-0.0326^{+0.0042}_{-0.0042}$ & $-0.0549^{+0.0050}_{-0.0050}$ \\
$b_2$  & $[0.8,1.2]$ & $0.9970713^{+0.0000065}_{-0.0000065}$ & $0.9969627^{+0.0000068}_{-0.0000070}$ \\
$m_2$  & $[-1,1]$ & $0.000690^{+0.000090}_{-0.000089}$ & $0.001894^{+0.000097}_{-0.000098}$ \\
$N_3$  & $[0.8,1.2]$ & $1.0600^{+0.0093}_{-0.0059}$ & $1.0710^{+0.0050}_{-0.0049}$ \\
Lon. Offset & ... & $33.04^{+0.70}_{-0.89}$ & $34.87^{+0.71}_{-0.78}$ \\
\enddata
 
\end{deluxetable*}

\subsection{Joint MIRI Eclipse Fits}
\label{sec:joint miri eclipse fits}

After  our individual eclipse fits, we performed \texttt{ThERESA} mapping on both MIRI eclipses simultaneously. We wanted to achieve the best possible constraints by using all available data and better resolve degeneracies between the ramp models by sharing a planet map across both eclipses. We used the \texttt{ThERESA} settings as described above and enforced the same astrophysical model across both eclipses, while allowing systematic ramp, normalization, and positional detrending parameters to vary between eclipses. 

\begin{figure}[ht!]
\centering
	\includegraphics[width=3.5in]{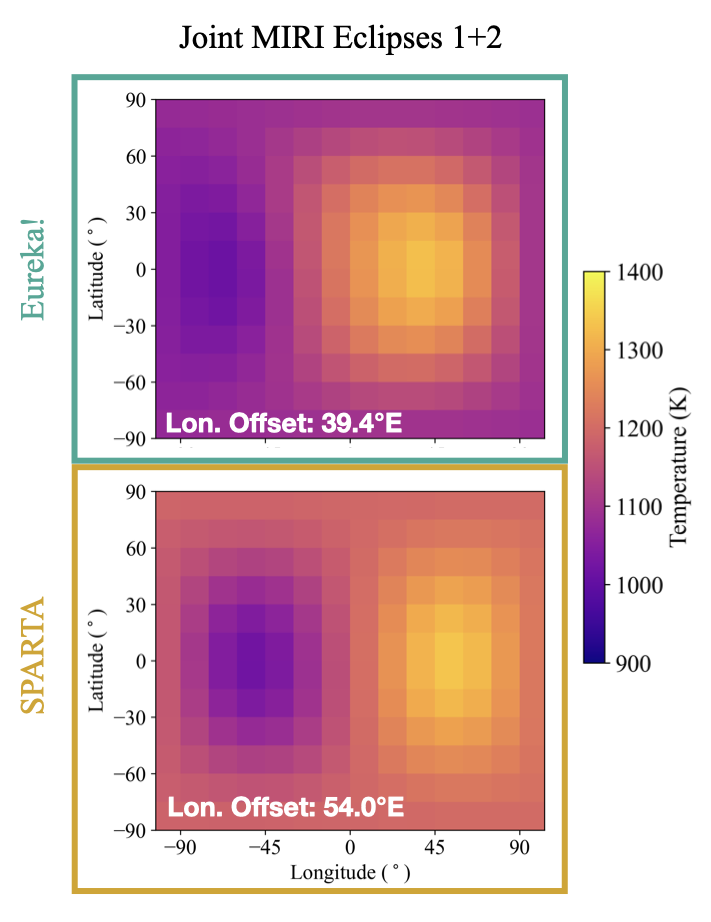}
    \caption{Dayside maps of HD189733b fit to both MIRI eclipses. The top row shows a map fit to the \texttt{Eureka!} reduction, while the bottom row shows a map fit to the \texttt{SPARTA} reduction. Shown in each map is the resulting longitudinal hotspot offset; although we achieve high-precision maps, they are still inconsistent across different data reduction methods. Reasons for the disagreement are further discussed in Section \ref{sec:joint miri eclipse fits}.
    }
    \label{fig:maps_joint_MIRI}
\end{figure}

We performed this analysis on both the \texttt{Eureka!} and \texttt{SPARTA} reductions, yielding 2 maps (shown in Figure \ref{fig:maps_joint_MIRI}). Our preferred models, priors, and posteriors are shown in Table \ref{table:individual_fits_table}. Inconsistencies in these results proved difficult to overcome: a degeneracy exists between the linear instrumental ramp and the phase variation, which provided a major obstacle to properly fitting the eclipse map. This issue became all the more clear when comparing the ramp posteriors from our two reductions: the instrumental slope profiles differed between the two reductions, likely responsible for the differences in our resulting maps. \utn{Even though the ramp model formulation used on the \texttt{Eureka!} and \texttt{SPARTA} reductions were identical, the best fit to the \texttt{SPARTA} data found a larger instrumental slope.} \ut{By fitting an instrumental slope which is too large, we may find a higher hotspot offset because less of the out-of-eclipse curve slope is attributed to real phase curve signal. This drives the model fit phase curve peak further away from the eclipse, and increases the resulting hotspot offset.} 

\subsection{Pre-Flattened Joint MIRI Eclipse Fits}
\label{sec:pre-flattened joint miri eclipse fits}

In order to \ut{understand} the degeneracy between the instrumental ramp and the shape of the phase curve, we first attempted to leverage in-eclipse data to constrain the linear ramp. Data within the eclipse (after ingress and before egress) does not include planet signal, so the instrumental ramp should be the only cause of variation in that section of data. We fit a line to the bottom of each eclipse and removed that slope (attributed to the instrumental ramp) from the entire eclipse lightcurve. However, this method \ut{did not lead to consistent results between the two data reductions.} \ut{We noticed that our linear slope was unable to flatten the in-eclipse data, especially near ingress and egress. We then fit slopes to regions of varying length ($<15$ min) just after ingress and just before egress; our retrieved map showed a dependence on the bin size of the data we considered in fitting a linear ramp, which led us to believe that red noise was contributing to the discrepancies between model fits to each reduction. In particular, we found that considering a short region (2 min) at the sub red-noise timescale just near ingress and egress resulted in agreement between maps between the two reductions.} This analysis increased our confidence that the difference between the two reductions was indeed due to the degeneracy between the instrumental ramp and the phase curve shape, with the steeper \texttt{SPARTA} instrumental slope resulting in an effectively flatter astrophysical phase curve model, resulting in an increased longitudinal hotspot offset in our maps. \ut{This test informed how we treated the data moving forward, but this data-flattening method was not used to achieve our final mapping result.}

\subsection{MIRI + Spitzer Eclipse and Partial Phase Curve Fits}
\label{sec:miri and spitzer eclipse and partial phase curve fits}

In order to ensure a more robust result, and one less dependent on very few data points of in-eclipse data near ingress and egress, we tried another method of breaking the degeneracy between instrumental ramp and phase curve variation. More baseline observing time would help to characterize the shape of the phase curve and therefore avoid misattributing systematic signal to astrophysical sources and vice versa. Due to the length of \thisplanet{}'s orbit and relatively large longitudinal hotspot offset, the phase variation is very nearly linear during our MIRI observation window, making it easily mistaken for the linear systematic. We utilize the Spitzer $8 \mu m$ partial phase curve \citep{knutson2007}, which includes the phase curve turnover,  to give us this extra baseline and disentangle the astrophysical and systematic linear trends near the eclipse. We also consider the Spitzer $8 \mu m$ eclipses in our analysis to ensure completeness and the most constrained possible result.

We run \texttt{ThERESA} using 3 separate observations: the first MIRI eclipse, the second MIRI eclipse, and the seven Spitzer IRAC eclipses and partial phase curve \citep{agol2010, knutson2007}, with all Spitzer observations phase folded into one dataset ("eclipse 3"). Our Spitzer data is identical to that used in previous mapping studies of \citet{rauscher2018} and \citet{challener2022}. As in our previous analysis, we run \texttt{ThERESA} in 2D mode, but we use 5,000,000 samples and a burn-in of 250,000 to ensure chain convergence with our expanded dataset. Our MCMC explores parameter space for eigencurve components up to the defined complexity limit ($l_{max}=10, N=3$). Note that we disallow higher-component models in this fit, while in previous iterations we had allowed up to $N=6$. This is due to a correlation issue we ran into with higher-component maps: even if BIC model selection prefers higher-component maps, the basis eigenmaps become overly correlated. We institute this $N_max$ limit as another model selection criteria according to \citet{mansfield2020}. As in previous iterations, we are again fitting for the ramp parameters $b,~p_1,~A$ and $\tau$, the positional detrending parameters $d_x,~d_y$ and $d_w$, and (with the exception of the first eclipse) the normalization factor $N_j$. The normalization factor is important especially when using the Spitzer dataset from years prior, in order to rescale the eclipses to accommodate stellar brightness variation over long timescales. Map latitude and longitude cell resolution is in this case set to an arbitrary $48$ by $96$. 

\begin{figure}[ht!]
\centering
    \includegraphics[width=0.43\textwidth]{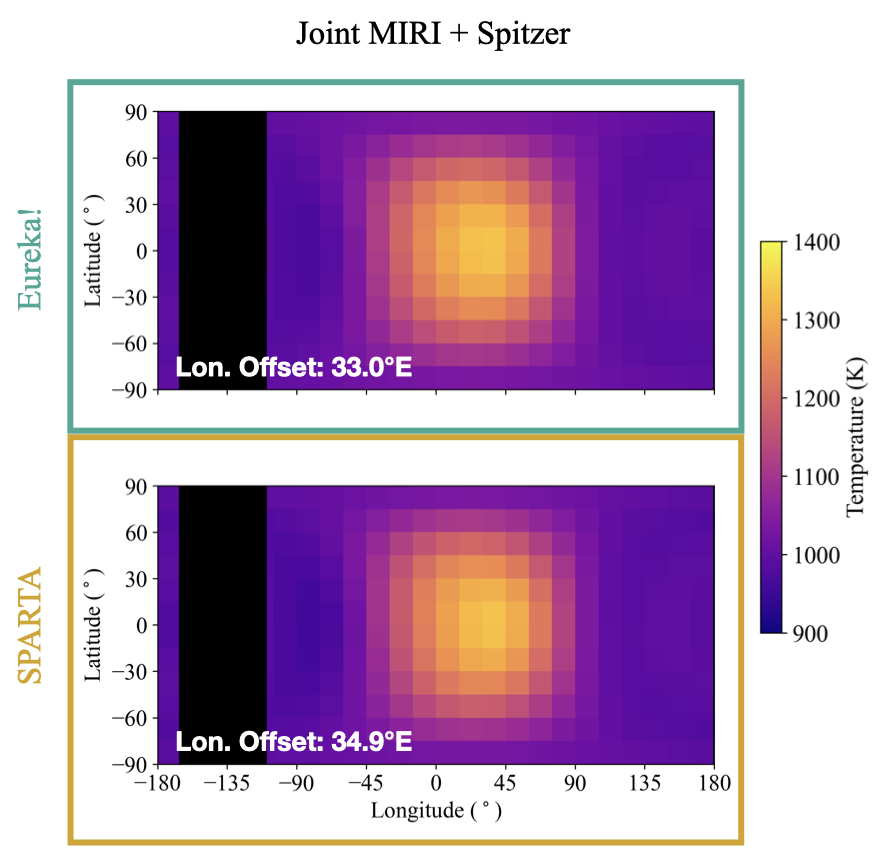}
    \includegraphics[width=0.43\textwidth]{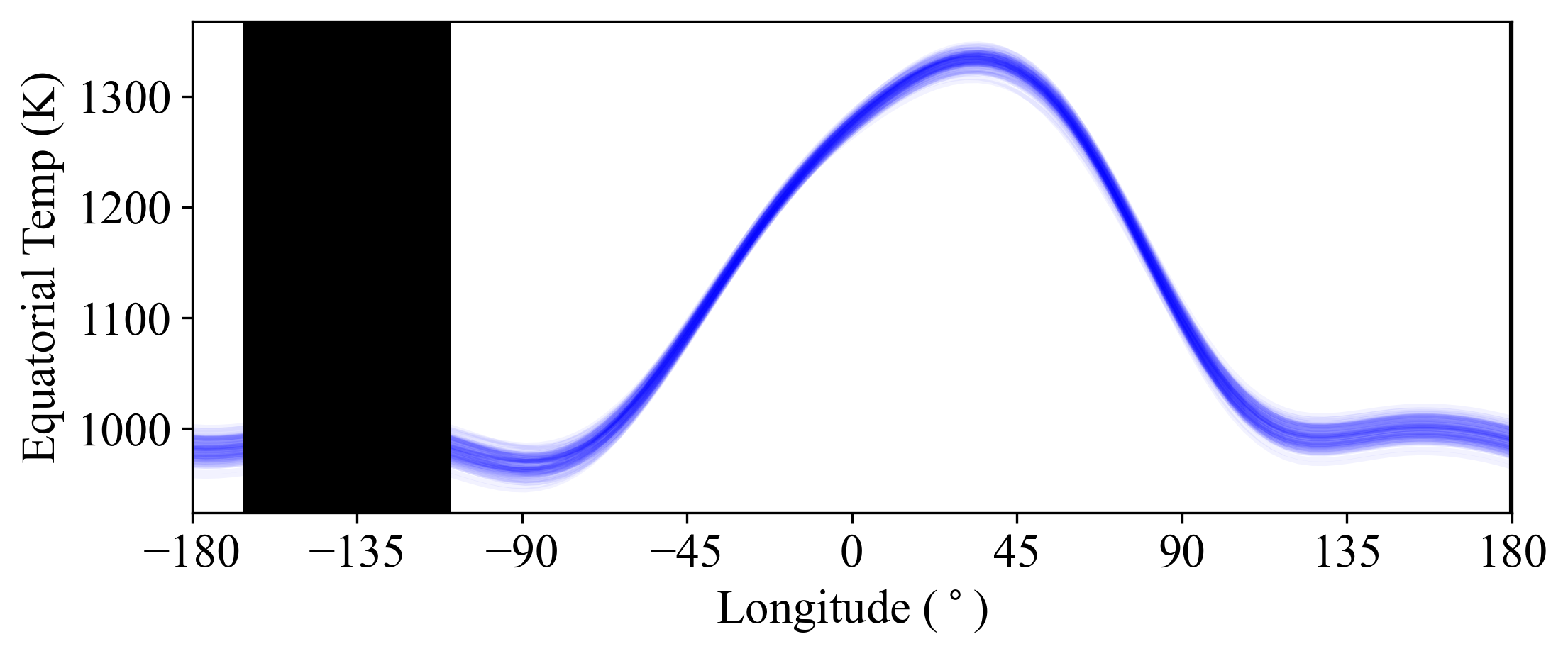}
    \includegraphics[width=0.43\textwidth]{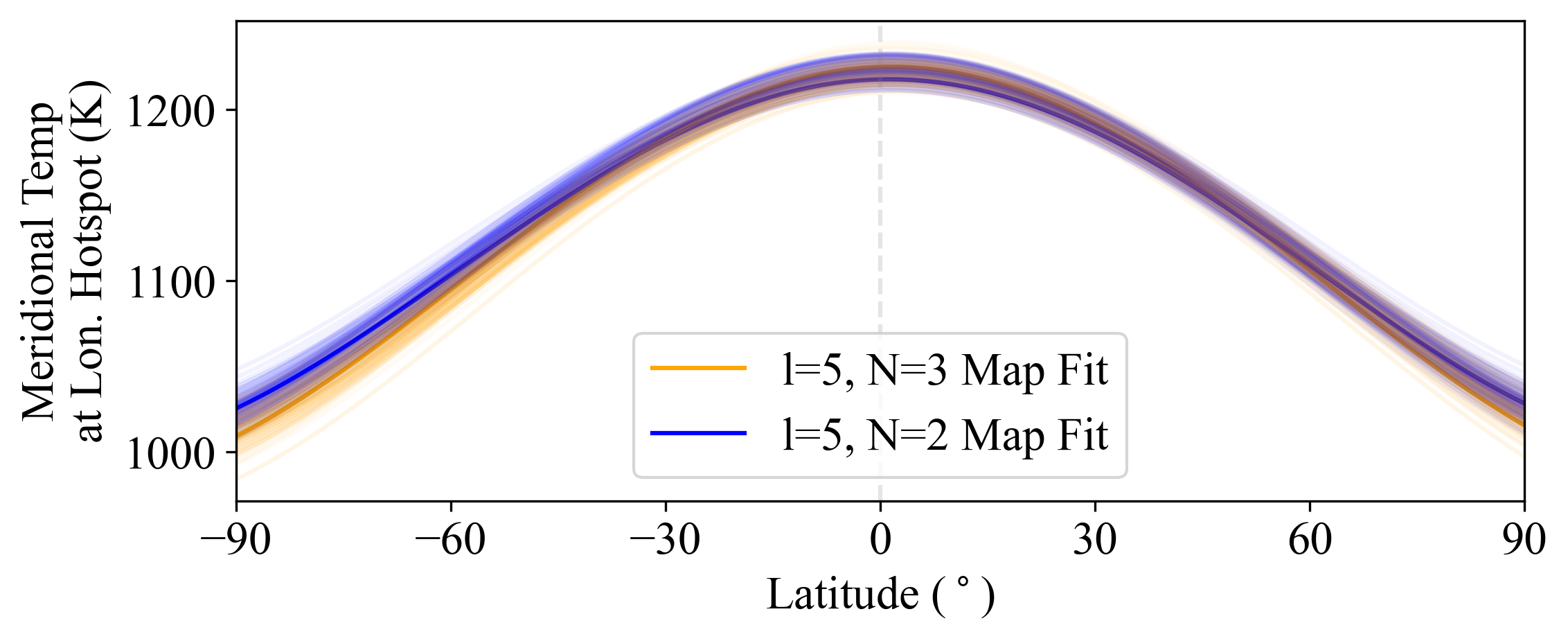}
    \vspace{-1pt}
    \caption{Dayside maps of \thisplanet{} fit to both MIRI eclipses, as well as seven Spitzer eclipses \citep{agol2010} and a Spitzer partial phase curve \citep{knutson2007}. The top row shows a map fit to the \texttt{Eureka!} reduction, while the second row shows a map fit to the \texttt{SPARTA} reduction. Shown in each map is the resulting longitudinal hotspot offset; unlike previously discussed analysis tests, these fits are in agreement with each other, due to the extended baseline from the partial phase curve breaking the degeneracy between the instrumental ramp and the phase curve shape. These maps also cover some of the nightside due to the added visible longitudes from the inclusion of the partial phase curve, although our analysis only focuses on the dayside. 
    \ut{The bottom two plots show 100 random draws from our posterior distribution at the equator vs. longitude, and the} \utn{hotspot meridian vs. latitude. The meridional temperature plot shows our overall best-fitting model ($l_{max}=5, N=2$) against the best-fitting 3-component fit ($l_{max}=5, N=3$, sensitive to latitudinal offsets). Based on the posterior distribution of our latitudinally-sensitive fit, we can confidently exclude a strong hemisphere-scale latitudinal offset ($>12^\circ$, by $3\sigma$).}}
    \label{fig:maps_joint_MIRI_Spitzer}
\end{figure}

The inclusion of the longer baseline worked as expected: the partial phase curve was able to break the degeneracy between the instrumental ramp and the phase curve, bringing our \texttt{Eureka!} and \texttt{SPARTA} reduction maps into agreement. The resulting maps are shown in Figure \ref{fig:maps_joint_MIRI_Spitzer} and as in previous cases, our preferred models, priors, and posteriors are shown in Table \ref{table:individual_fits_table}.

\ut{We note that combining the MIRI and Spitzer eclipses in this analysis assumes that the dayside map is not detectably time variable over the time period of our observations. Such variability has yet to be detected \cite[e.g.,][]{lally&vanderburg2022, murphy2023, li&shporer2024, wang&espinoza2024}, but modeling work has suggested the possibility of hotspot offset and temperature variability. \citet{komacek2020} predicts variations of less than $1\%$ (less than $0.5\%$ in most cases, corresponding to approx. <10K) in the planet’s dayside temperature, which is well below our capability of detection for a single eclipse observation. Our temperature map uncertainties on single eclipses ($\approx 30$ K) are not sensitive to this level of variability. Furthermore, \citet{inglis2024} shows agreement in the planet’s brightness at the two epochs observed by JWST, and \citet{kilpatrick2020} shows no variability in Spitzer observations of HD189733b spanning 5 years. 
Predictions of hotspot offset variability depend on assumptions about drag timescales, but, for planets like HD189733b, are not expected to exceed 3 degrees. This is similarly beneath our detection threshold for single-eclipse maps.
}

\section{Results} 
\label{sec:results}

\begin{figure*}[ht!]
\centering
	\includegraphics[width=\textwidth]{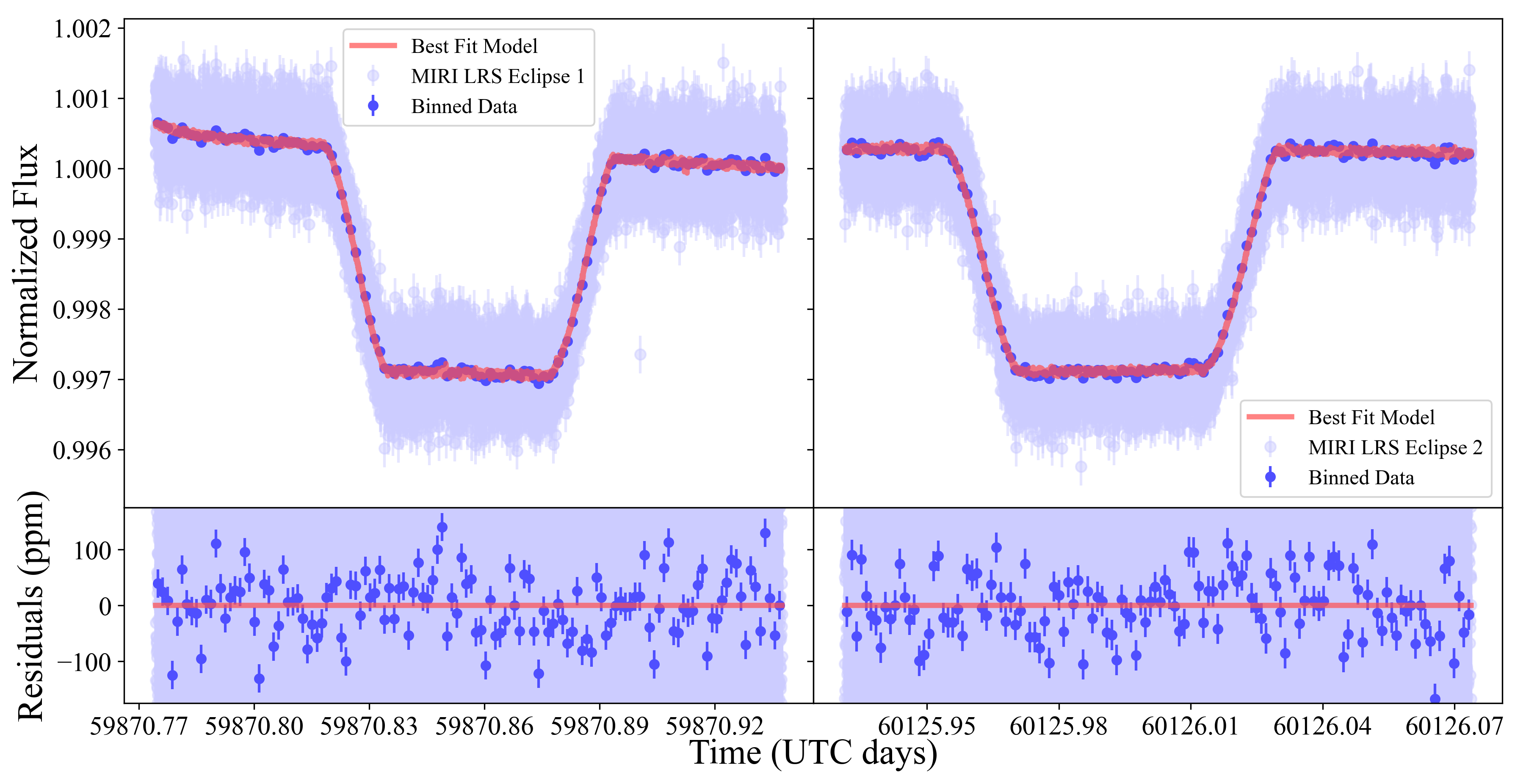}
    \caption{Lightcurves of the two MIRI LRS eclipses analyzed in this work. The red best-fit model shown over both eclipses is the result of a simultaneous ThERESA fit of both MIRI LRS eclipses and, although not shown on the plot, the seven Spitzer IRAC $8\mu m$ eclipses of \citet{agol2010} and the partial phase curve of \citet{knutson2007}. The MIRI LRS data are shown unbinned, as well as binned to 130 points (about 2 minutes) for visual clarity. The joint fit of the 8 eclipses and partial phase curve achieves an SDNR of 334 ppm, which is $1.8 \times$ the photon noise floor of 181 ppm.}
    \label{fig:eclipses}
\end{figure*}

\begin{figure*}[hbt!]
\centering
	\includegraphics[width=6.5in]{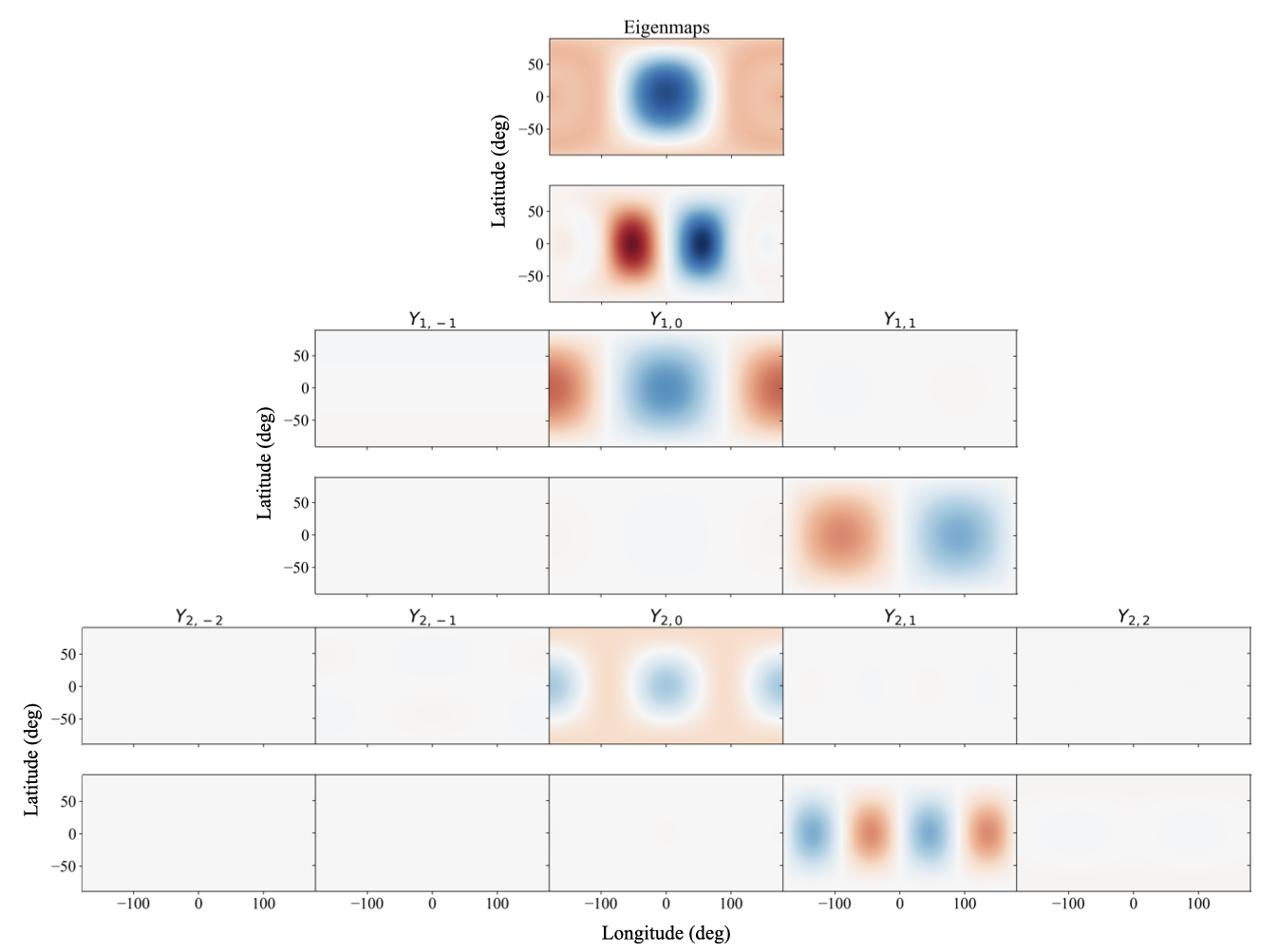}
    \caption{Schematic showing the weighted eigenmap components incorporated into our joint best-fit eclipse map shown in the \texttt{Eureka!} row of Figure \ref{fig:maps_joint_MIRI_Spitzer}. \textbf{The top plot} shows the two mapping components, the weights of which are described by $C_1$ and $C_2$ in Table \ref{table:individual_fits_table}. Our resulting map is a linear combination of these two components, in addition to a uniform brightness component. \textbf{The rows beneath} show their constituent harmonic components, with the strength of each component corresponding to its opacity. Some contributing higher-degree spherical harmonic modes (3rd, 4th and 5th) are not shown in this plot, due to their contribution being minor compared to the modes shown.}
    \label{fig:eigenmaps}
\end{figure*}

\begin{figure}[hbt!]
    \includegraphics[width=0.48\textwidth]{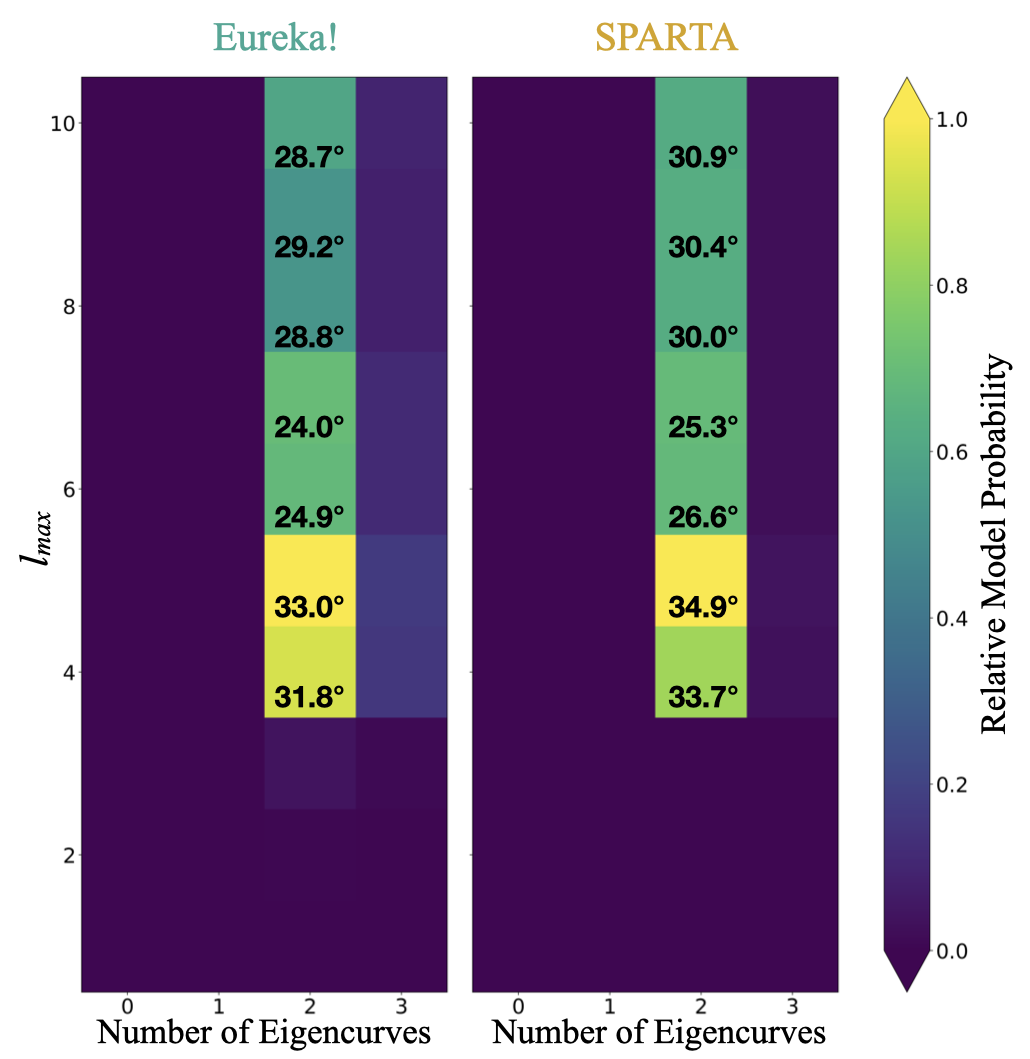} 
    \caption{A visualization of our model selection process. Relative Model Probability (RMP) is related to the difference in BIC values between each possible fit ($RMP = e^{(-\Delta BIC/2)}$). Each square is a possible fit explored by \texttt{ThERESA}. Resulting longitudinal offsets are shown within each square for a selection of relatively probable models.}
    \label{fig:bics}
\end{figure}

We present the results of our analysis in Figure \ref{fig:eclipses}, which shows the lightcurves of the two eclipses used in this analysis, with the best-fit curve from the joint fit overplotted. 
Our joint fit map uses the two highest-variance eigencurves ($N=2$) computed from 5th degree spherical harmonics ($l_{max}=5$), as visualized in Figure \ref{fig:eigenmaps}. Inclusion of up to the 5th harmonic degree is supported by our BIC values: $l_{max}=5$ is weakly preferred to $l_{max}=4$ ($\Delta BIC<1$), and strongly preferred to $l_{max}=3, 2$ and $1$ ($\Delta BIC>6$). See Figure \ref{fig:bics} for more model selection details.  

\begin{figure*}[hbt!]
    \centering
    \includegraphics[width=0.96\textwidth]{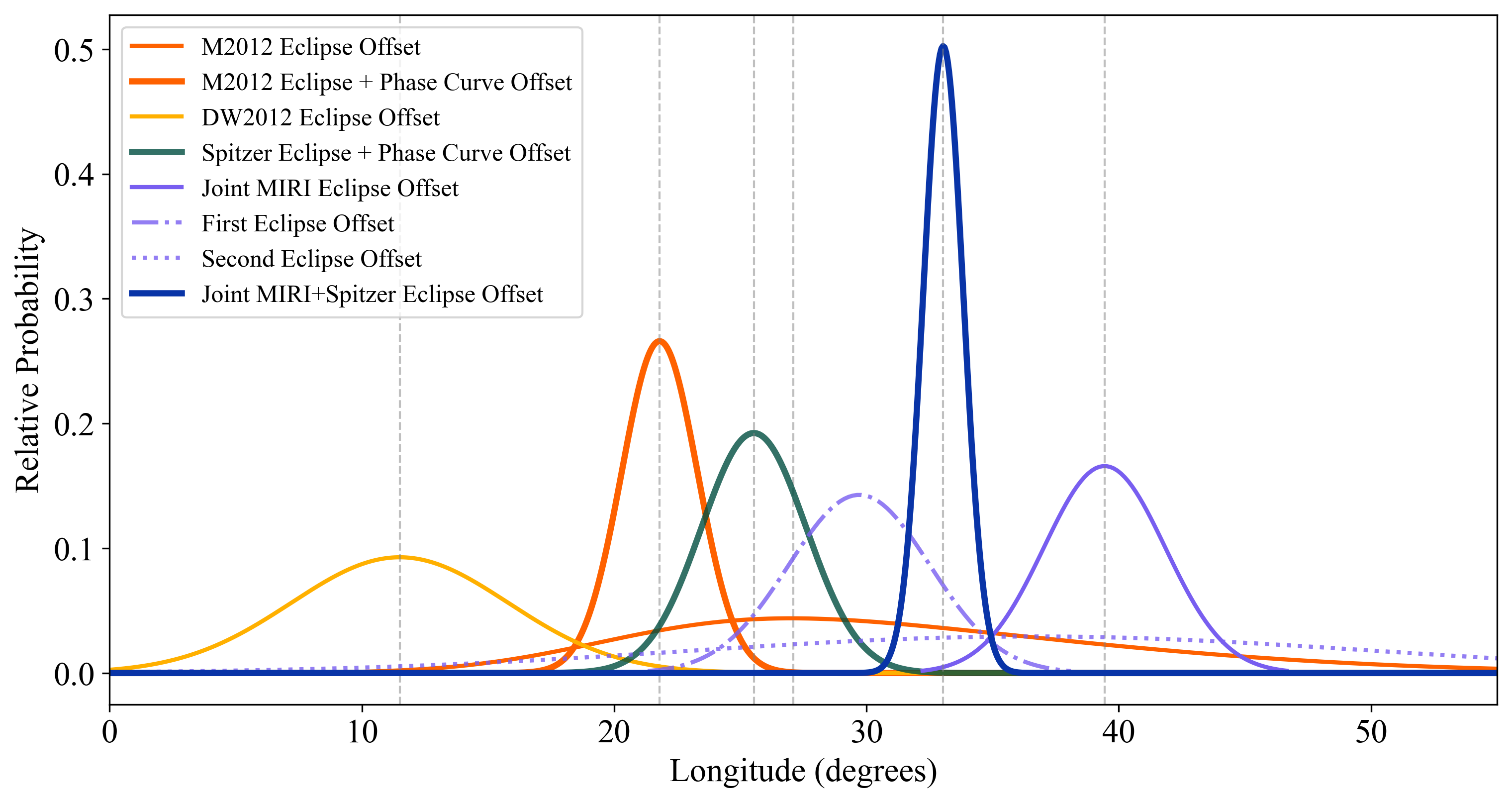}
    \includegraphics[width=0.835\textwidth]{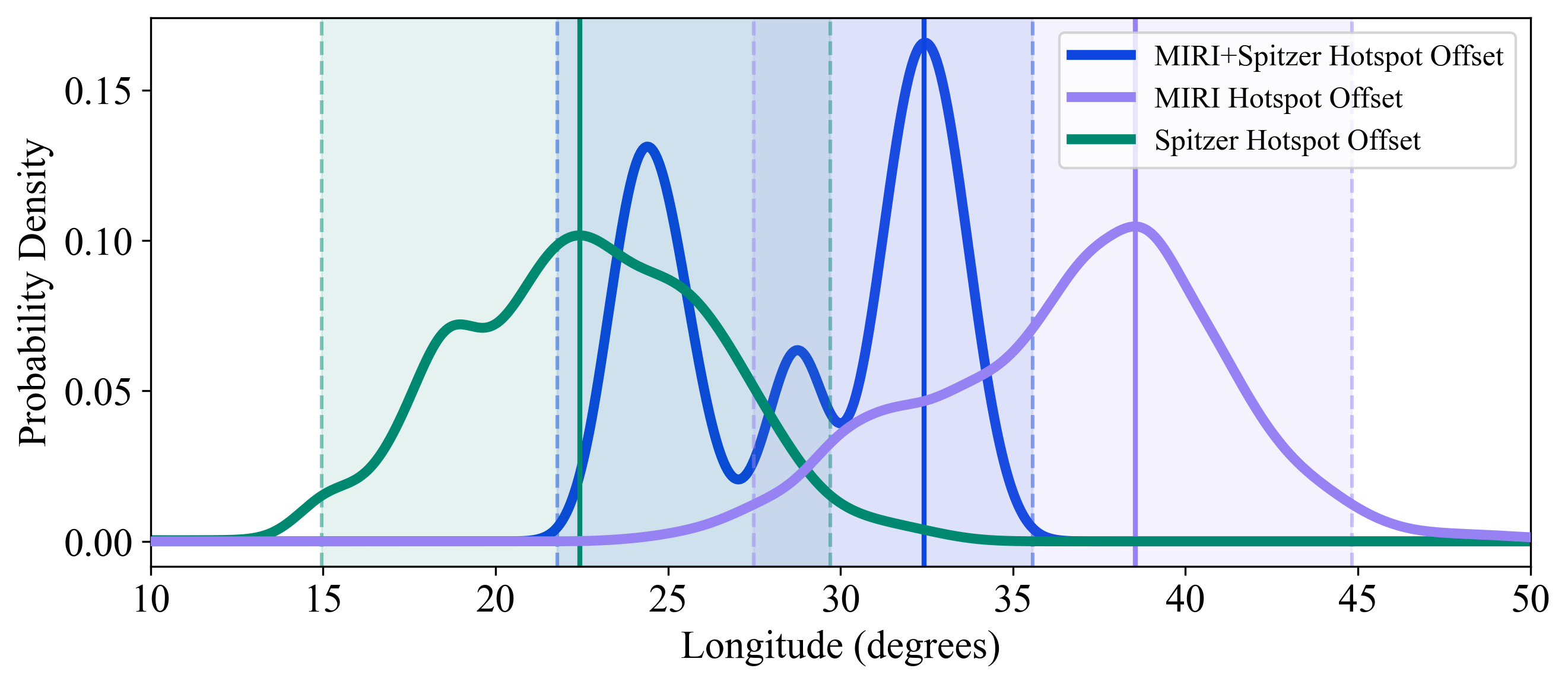}
    \caption{\utnn{\textbf{Top:}} Distributions of best-fit hotspot offsets from the MIRI joint fit (purple), MIRI+Spitzer joint fit (blue), and the individual fits presented in this paper (purple dotted and dot-dash). Best-fit results are shown for the fits presented in this paper; the joint fit displays the distribution from an $l=5, N=2$ fit, the first individual eclipse fit shows the distribution from an $l=3, N=2$ fit, and the second eclipse from an $l=1, N=3$ fit (as displayed in the best-fit maps shown throughout the paper, as well as Table \ref{table:individual_fits_table}.). Also shown are the results of previous eclipse mapping studies \citet{majeau2012} \textbf{(M2012)} and \citet{dewit2012} \textbf{(DW2012)}, as well as a test fit using only Spitzer eclipse and phase curve data ($l=5, N=2$; green). Single-model offsets presented in this paper are consistent with the offset presented from a fit to the Spitzer eclipses in M2012, but about $7 \sigma$ larger than the Spitzer eclipse + phase curve result from M2012, and $5 \sigma$ larger than the result presented in DW2012. The MIRI eclipses provide improved signal and constraints over what a Spitzer phase curve alone could provide. We note that the Spitzer-only and MIRI-only hotspot offsets have a ~$3\sigma$ separation. This is likely because the eigenmap calculation process is dependent on the time sampling of the lightcurves: \utnn{while each of these show the same $l=5, N=2$ model, there are differences in the two highest-variance eigenmaps used to form the model basis, contributing to differences in the resulting hotspot offsets. As discussed in Section \ref{sec:results}, hotspot offset can vary with model assumptions; single model fits, even if they are all $l=5, N=2$ models, have some discrepancies.} \utnn{ The degeneracy present in the MIRI-only dataset also has a significant impact on its longitudinal hotspot offset, complicating its use as a point of comparison.} \utnn{\textbf{Bottom:}} All \texttt{Eureka!} models $N=1$ to $N=3$ are weighted by relative model probability \utnn{(as described in Figure \ref{fig:bics})}, and their posteriors are stacked to create the distribution shown. \utnn{This process is done for fits to the MIRI dataset, the Spitzer dataset, and the combined MIRI+Spitzer dataset. Credible regions for each fit are shown by the highlighted regions.}}
    \label{fig:offsets}
\end{figure*}

Our posteriors for each fit appear in Table \ref{table:individual_fits_table}. Our preferred joint fit finds an eastward hotspot offset of $33.0^{+0.7}_{-0.9}$ degrees. The longitudinal offsets resulting from several of our attempted fits, \utnn{as well as previously published fits,} are shown in Figure \ref{fig:offsets}.

\utnn{The top panel of Figure \ref{fig:offsets} provides an apt comparison to previous work by presenting a result and uncertainty based on a single best-fit model.} While our best-fit model is very well-constrained compared to previous studies, we attempt to address the model-dependence discussed in this and previous eclipse mapping studies by folding together the posteriors of several relatively probable models. We consider models from $l_{max}=1 - 10$ and $N=1 - 3$. We then use the Bayes factor of each possible model relative to the best-fit (minimum BIC) model, which we use to weight each model by its likelihood and create a stacked posterior distribution (shown in the bottom panel of Figure \ref{fig:offsets}). \utnn{This process emphasized the importance of accommodating model dependence into the uncertainty on our hotspot location; as shown in the top panel of Figure \ref{fig:offsets}, the single-model fit to the Spitzer dataset appears inconsistent with the single-model fit to the combined Spitzer and MIRI datasets. However, when model dependence is considered (as in the bottom panel of Figure \ref{fig:offsets}), the Spitzer-only hotspot offset is within 1$\sigma$ of the MIRI+Spitzer hotspot offset.}

Formally our 2D \texttt{ThERESA} model also fit for latitudinal offsets, finding an offset of $2.2^{0.1}_{-0.1}$ degrees. Realistically, however, the latitudinal offset is unconstrained, since these 2-component models lack the flexibility to verify latitudinal temperature structure \citep{challener2024}. \utn{Although it is not preferred by our BIC minimization,} a 3-component model which does allow for flexibility in latitudinal structure finds a latitudinal hotspot offset of $6.1\pm1.6$ degrees, \utn{with a $3\sigma$ constraint to $<12^\circ$ away from the equator.} \utn{This excludes a large hemisphere-scale latitudinal asymmetry, but allows for the possibility of a small latitudinal offset, providing a promising avenue for follow-up observation.} 

In order to ensure the robustness of our result, we tested if our longitudinal offset would vary significantly if planetary input parameters were varied within $1\sigma$ \citep{coulombe2023, challener2024}. We tested variations of semi-major axis $a$, planet radius $R_p$, eclipse time $t_e$, \ut{and inclination $i$}. Variations of $1\sigma$ in $a$ did not change the resulting longitudinal offset. Similar variations in $R_p$ and $t_e$ each changed the hotspot offset by $<0.1^\circ$, which is consistent with the original result. \ut{Inclination had the largest effect: a $1 \sigma$ increase in inclination decreased the hotspot offset by $1.2^\circ$, while a decrease in inclination had a $<0.5^\circ$ effect. This is not unexpected; orbital inclination has been shown in previous works to have a relatively strong effect on the eclipse mapping signal \citep{majeau2012, boone2024, hammond2024}. That said, this change is still relatively small, within $<1\sigma$ of our result}.  While noting that the variations in our input parameters had mostly insignificant affects on our posteriors, we do note that our fit value uncertainties may be slightly underestimated, since eigenmapping requires fixing orbital parameters. 

\section{Discussion}
\label{sec:discussion}

Our analysis tests and their results leave us with several important notes on approaching MIRI LRS eclipse mapping, now and in the future. Mainly, we find that a significant degeneracy exists between the MIRI LRS instrumental ramp and the phase curve variation. This degeneracy must be treated in order to produce accurate eclipse maps. This can be done by leveraging in-eclipse data to constrain the linear slope of the instrumental ramp, but the presence of red noise may complicate this approach. A more reliable method to break this degeneracy is to use longer baseline observations. This helps in two ways: firstly, the instrumental ramp damps out over time, so obtaining longer baselines can help ensure that the instrumental ramp is minimal by the time the eclipse occurs. Secondly, longer baseline observations will include more of the phase curve, which can help avoid attributing true phase curve variation to instrumental ramp and vice versa. 

\subsection{Comparison with Previous Eclipse Mapping Studies}
Our joint-fit result of an eastward hotspot offset of $33.0^{+0.7}_{-0.9}$ is consistent with the eclipse-only mapping results from \citet{majeau2012}, who favored an eastward offset of $27.1^{+43.5}_{-9.0}$ degrees. However, our result is significantly separated from that of \citet{dewit2012}, who found an eastward offset of $11.5 \pm 4.3$ degrees. The difference between these results may be due to several different analysis approaches: \citet{dewit2012} opted to leave more planetary parameters unfixed, including impact parameter, eclipse mid-time, eccentricity, and stellar density (parameters which have since been better constrained in follow-up studies); the paper explored the correlation between those parameters, and how changing them altered their retrieved eclipse map. (Although our analysis fixes a zero eccentricity, we effectively explore small changes in eccentricity in the robustness test described at the end of Section \ref{sec:results}; we change the eclipse time by $1\sigma$, which does not significantly impact our resulting hotspot offset). In addition, \citet{dewit2012} emphasizes that, due to the model dependence of their results, they are hesitant to definitively summarize their result as one longitudinal hotspot offset value, as this sheds context; instead, they prefer a more extensive discussion of the brightness distribution. We encountered similar difficulties with model dependence, especially before considering the longer baseline phase curve observation provided from Spitzer \citep{knutson2007}. We additionally address the model dependence by stacking the posterior distributions of several relatively probable models, shown in Figure \ref{fig:bics}, to find an eastward offset of $32.5^{+3.0}_{-10.6}$ \utnn{(see Figure \ref{fig:offsets})}. This is not a significantly different offset than our result from only our best-fit model; however, it does significantly expand our credible region from $32.1 - 33.7$ to $21.9 - 35.5$, which better reflects the uncertainties in the modeling process. 

For comparison’s sake, we tested our joint-fit results when limited to a first-order harmonic fit \citep[as used by][]{majeau2012} and found an offset of $28.1\pm1.4$ degrees eastward. This is consistent with the eclipse-only first-order results from \citet{majeau2012} of $27.1^{+43.5}_{-9.0}$ degrees (and far more constrained), and is within $4\sigma$ of their eclipse + phase curve result of $21.8\pm1.5$ degrees. We then tested a second-order harmonic fit ($l=2, N=2$, as used by \citet{challener2022}) and found an offset of $24.9\pm1.6$ degrees, which is within $2\sigma$ of the \citet{challener2022} result of $21.8\pm1.5$ degrees. Future discussion of our result refers to the best-fit offset achieved with the 5th degree 2-component map, as this is the model preferred by our BIC minimization comparison. To the point made in \citet{dewit2012} regarding the importance of the shape of the distribution, some of our lower-harmonic fits came up with very similar longitudinal offset values as our 5th order map, but the higher-order models allowing a steeper drop-off in temperature eastward of the hotspot were better fits to our data. 

Our current data quality did not justify the use of more than 2 to 3 mapping components, even with our lower-degree fits. When we only use the first two components of each harmonic degree, we are only sensitive to relatively large-scale asymmetries and temperature gradients. Our data support an eastward offset which drops off steeply eastward of the hotspot. With our current data, we cannot be confident in any smaller trends fit by the model, but our results may indicate the presence of a non-zero latitudinal offset--more definite results require follow-up observations. \ut{Although we do not detect a latitudinal offset with any certainty, the observations and analysis have the capability to detect and constrain latitudinal offsets; our non-detection here rules out any extreme latitudinal asymmetry, as, if a highly latitudinally offset hotspot was present, our BIC minimization algorithm would prefer a higher-component model.}

\begin{figure*}[ht!]
\centering
   \includegraphics[width=6.5in]{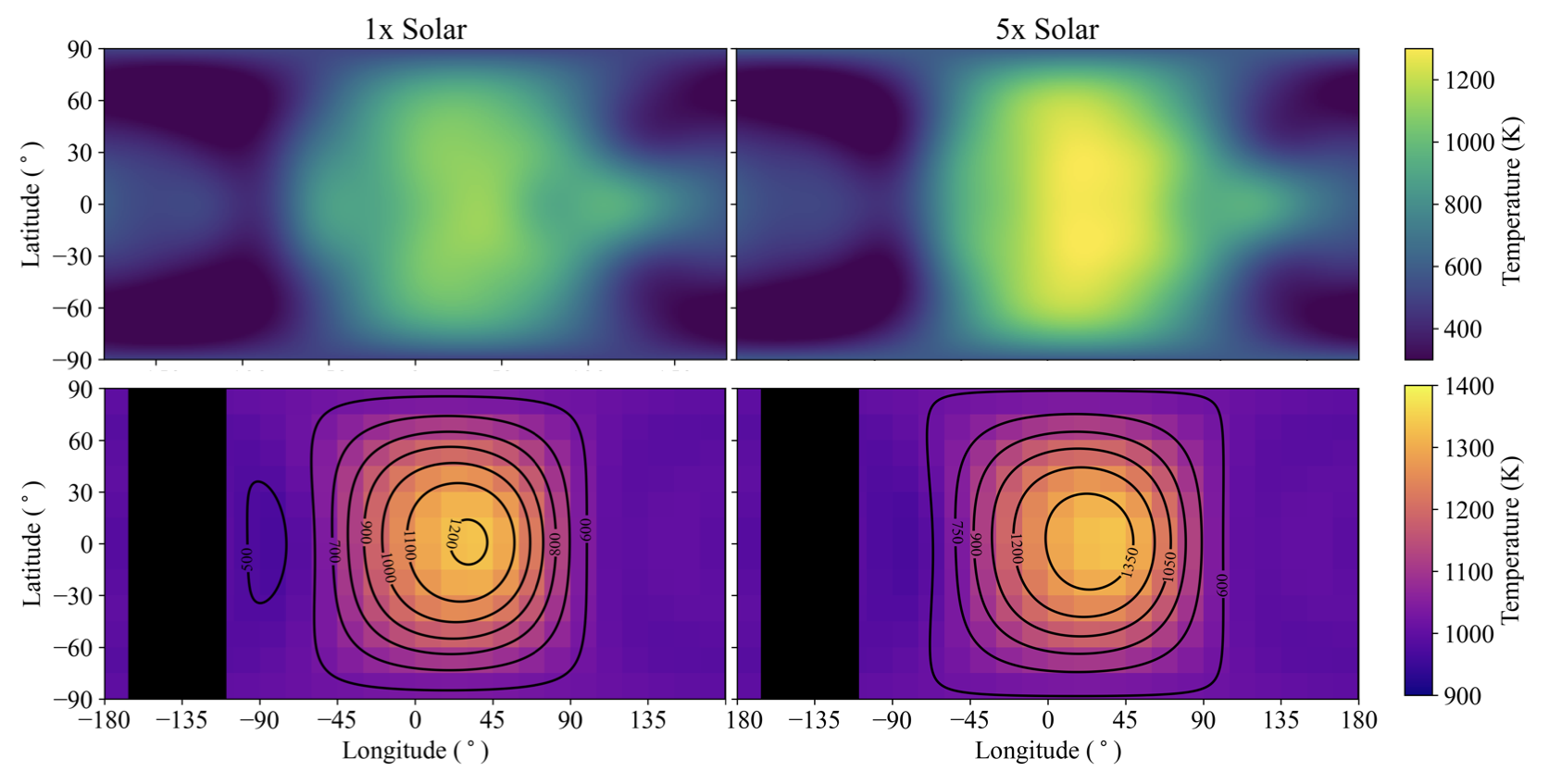}
   \includegraphics[width=6in]{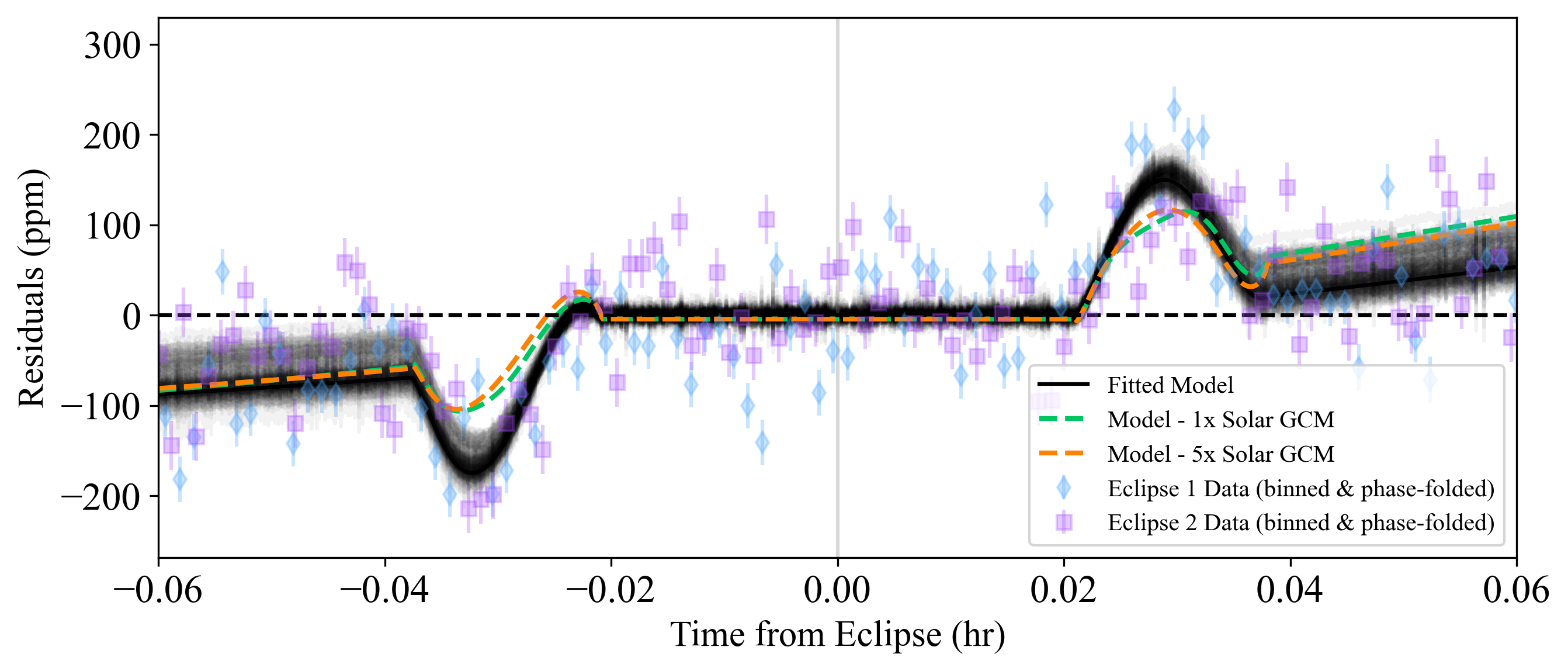}
   \caption{\textbf{The top row} of this figure shows two GCM models at $8 \mu m$; the left column shows a model assuming 1$\times~$ solar metallicity, and the right column assumes 5$\times~$ solar metallicity. \textbf{The middle row} shows our retrieved dayside joint-fit map (the same map shown in the \texttt{Eureka!} row of Figure \ref{fig:maps_joint_MIRI_Spitzer}), overlaid with contour maps showing an eigenmap representation of the respective GCM maps shown in the top row. \textbf{The bottom row} shows residual differences between a uniform dayside model (horizontal dotted line) and our best-fit eclipse model, and each of the two GCMs. Data from both eclipses are binned down to a 2 minute cadence for visibility, \ut{and are consistent to a $~1.3\sigma$ level.}
   }
   \label{fig:GCM}
\end{figure*}

\subsection{Comparison with GCMs}
Our result is also largely consistent with predictions from GCMs of HD~189733b \cite[e.g.,][]{showman2009,kataria2016, flowers2019}. Such models have shown that global advection strongly affects heat transport, such that we should expect a noticeably eastward-shifted hotspot in HD~189733b's atmosphere. Wind and temperature profiles of this planet at the $\sim$100 mbar level (a pressure near the IR photosphere probed in the $8 \mu m$ MIRI bandpass) suggest an eastward equatorial jet, shifting the hottest regions $\sim$50$^\circ$ east \citep[e.g.,][]{flowers2019}.

To quantitatively compare the GCMs with the data and our retrieved maps, we used \texttt{starry} to compute 7th-order spherical-harmonic representations of the GCM $8 \mu m$ emission maps. We then extracted the components of the GCM maps that we would expect to retrieve using our optimal joint-fit model ($l_{max}=5, N=2$) -- in other words, an eigenmap representation of the GCM maps that can be directly compared against our fitted maps (see \citealp{challener_rauscher2023, hammond2024} for more detailed descriptions). We also used the spherical-harmonic representations of the GCM maps to calculate the expected eclipse light curves from the GCMs for comparison against our observations. We test versions of the $8 \mu m$ emission GCMs which assume a metallicity of 1$\times~$ solar and 5$\times~$ solar. These direct comparisons, shown in Figure \ref{fig:GCM}, underscore the agreement between GCM predictions and our retrieved maps. 
The GCM and best-fit map are largely consistent in structure, with the hottest regions of the planet well-correlated in longitude and latitude, and the same steep drop-off east of the hotspot. However, the temperatures derived from the GCM appear slightly cooler than the temperatures derived from the best-fit map. This could be attributed to differences in atmospheric composition and internal heating, which can act to elevate atmospheric temperature \citep[e.g.,][]{lewis2010, kataria2015}. The inclusion of hazes \citep{steinrueck2023} or a higher metallicity \citep{kataria2015} could both serve to enhance the day-night temperature contrast on HD~189733b. However, it is unclear to whether the inclusion of hazes may reduce the hotspot offset (and to what degree), which may adversely affect the agreement between GCM and best-fit map. Future work will investigate these effects when comparing GCMs to MIRI spectroscopic eclipse maps. Nevertheless, these comparisons suggest good agreement between the GCM and the observed maps, without the need to invoke effects due to atmospheric drag as was necessary in some previous analyses \citep{coulombe2023, challener2024}. 

\section{Conclusion}
\label{sec:conclusion}

This work uses JWST MIRI eclipse observations of much higher SNR than was previously available, paired with previous Spitzer IRAC $8 \mu m$ eclipses and a partial phase curve, to create a 2D $8 \mu m$ dayside temperature map of \thisplanet{}. In particular, we use a 2-component 5th degree harmonic fit to the MIRI and Spitzer data to find an eastward hotspot offset of $33.0^{+0.7}_{-0.9}~$. The improvement in data quality offered by JWST allows for a significantly more constrained measurement of the location of the peak brightness temperature: previous eclipse mapping work using Spitzer IRAC observations of seven eclipses had an uncertainty range 2 to 11 times wider than that of our MIRI-eclipse-only result ($39.4^{+2.3}_{-2.5}$), which uses only 2 eclipses \citep{majeau2012, dewit2012}. This level of constraint, from about a quarter of the observing time, demonstrates the power of JWST MIRI observations in comparison to previous telescopes. 

\utnn{That being said, this work also emphasizes the importance of incorporating model-dependence into uncertainty estimates for hotspot offsets; when our best-fit model is computed using a weighted average based on Bayes factor (see Figures \ref{fig:bics} and \ref{fig:offsets}), we find an eastward hotspot offset of $32.5_{-10.6}^{+3.0}$, which is considerably less constrained than our single-model approach. Rather than suggesting that new JWST observations provide no improvement over previous datasets, this suggests that previous studies were underestimating uncertainties on their reported hotspot locations by presenting only a single best-fit model. Despite the model dependency, the eclipse maps presented in this work represent a great improvement in our understanding of \thisplanet{}, including latitudinally sensitive eclipse maps, effectively constraining the hotspot offset to within ~12$^\circ$ of the equator. This work also promises a robust future for eclipse mapping of other hot Jupiter exoplanets.} JWST MIRI data quality, with appropriate baseline to treat degeneracies, will allow for the type of eclipse-mapping analysis carried out in this paper to be applied to many more planets than was previously possible, and with less data than was previously required \citep{boone2024}. In addition, the generalizable instrumental ramp fitting capabilities of \texttt{ThERESA} developed for this paper's analysis will ease future eclipse mapping work and allow flexible analysis of multiple eclipses and/or multiple bandpasses. 

Future proposals intending to carry out similar eclipse mapping studies should recognize the utility of a long baseline of data prior to the eclipse. Degeneracies between instrumental ramps and astrophysical signals were a challenge in our analysis; this could be minimized by ensuring that the strongest ramp signal occurs earlier than the eclipse, and does not significantly overlap (especially with ingress and egress). Longer baseline observations also provide more phase variation signal to help break the phase/ramp degeneracy; if the baseline includes the overturn (or the local maximum) of the phase curve, that is particularly beneficial in providing a non-linear signal to disentangle with the linear component of the instrumental ramp. 

As discussed in the introduction, \thisplanet{} is much more than a test case for future planets--according to the eclipse mapping metric posed in \citet{boone2024}, it is the most favorable planet for MIRI LRS eclipse mapping, and much more work can be done to study its dayside. Although our current data is unable to distinguish real latitudinal offsets, fits which incorporate a non-zero latitudinal offset are more favored than uniform dayside fits. If latitudinal asymmetries are in fact present, further MIRI observations of \thisplanet{} eclipses are likely to reveal them. In addition, the observations used in this work are spectroscopic and extend beyond the wavelength range analyzed here. In future work, a wavelength-dependent mapping analysis \ut{may additionally yield pressure-dependent information about the planet's dayside atmosphere}. 

\begin{acknowledgments}
This material is based upon work supported by the National Science Foundation Graduate Research Fellowship under Grant No. DGE – 2139899. This work is based on observations made with the NASA/ESA/CSA James Webb Space Telescope. The data were obtained from the Mikulski Archive for Space Telescopes (MAST) at the Space Telescope Science Institute, which is operated by the Association of Universities for Research in Astronomy, Inc., under NASA contract NAS 5-03127 for JWST. These observations are associated with the program JWST-GO-2021 (PI Kilpatrick). Support for program JWST-GO-2021 was provided by NASA through a grant from the Space Telescope Science Institute, which is operated by the Association of Universities for Research in Astronomy, Inc., under NASA contract NAS 5-03127. This research has made use of NASA's Astrophysics Data System and the NASA Exoplanet Archive, which is operated by the California Institute of Technology, under contract with the National Aeronautics and Space Administration under the Exoplanet Exploration Program.
We thank E. Agol and N. Cowan for the Spitzer eclipse and phase curve data used in this paper. \utnn{We thank the referee for their constructive feedback, which strengthened this paper.}
Data products from this paper will be available in this Zenodo repository: TBD.

\end{acknowledgments}

%

\vspace{5mm}
\facilities{JWST MIRI, Spitzer IRAC}
\software{numpy \citep{np}, matplotlib \citep{plt}, astropy \citep{astropy2018}, scipy \citep{2020SciPy-NMeth}, ThERESA \citep{challener2022}, mc3 \citep{mc3}, starry \citep{starry}}




\bibliography{ref}{}
\bibliographystyle{aasjournal}

\end{document}